\RequirePackage{etoolbox} 

\newtoggle{refereever}

\nottoggle{refereever}{
    \documentclass[bibyear]{aa}  
}{
    \documentclass[referee, bibyear]{aa}  
}

\usepackage[varg]{txfonts}

\usepackage[utf8]{inputenc}
\usepackage[T1]{fontenc}
\usepackage[english]{babel}

\usepackage{natbib}
\bibpunct{(}{)}{;}{a}{}{,}  \usepackage{csquotes}
\nocite{*}

\let\cite\biblatexcite
\newcommand*{\cite}{\citealp}
\newcommand*{\textcite}{\citet}

\usepackage{amsmath}
\usepackage{amsfonts}
\usepackage{amssymb}
\usepackage{graphicx}

\usepackage{float}

\usepackage{diagbox}
\usepackage{booktabs}
\usepackage{multirow}
\usepackage{multicol}
\usepackage{makecell}
\usepackage{tabularx}
\usepackage{adjustbox}
\usepackage{placeins}[section]
\usepackage[separate-uncertainty=true, separate-uncertainty-units = single, range-phrase=-, range-units=single]{siunitx}

\nottoggle{refereever}{
    \usepackage[table]{xcolor}
    
    \definecolor{lightgreen}{RGB}{144,238,144}
    \definecolor{lightcoral}{RGB}{240,128,128}
    \definecolor{darkgreen}{RGB}{0,100,0}
    
}{}

\usepackage[
        hidelinks, 
        colorlinks=true,
allcolors=blue,
        allbordercolors=blue,
        linktoc=all,
    ]{hyperref}

\usepackage{lscape} \usepackage{rotating}

\usepackage{mathtools}

\title{The Repeating Flaring Activity of Blazar \object{AO 0235+164}}

\author{
	         Juan Escudero Pedrosa\thanks{\email{jescudero@iaa.es}}\inst{,\ref{affil:IAA-CSIC}}
	    \and Iván Agudo\inst{\ref{affil:IAA-CSIC}}
	    \and Andrea Tramacere\inst{\ref{affil:UniGe-astro}}
	    \and Alan P. Marscher\inst{\ref{affil:BU-blazar}}
	    \and Svetlana Jorstad\inst{\ref{affil:BU-blazar},\ref{affil:SPBU}}
	    \and Z. R. Weaver\inst{\ref{affil:BU-blazar}}
	    \and Carolina Casadio\inst{\ref{affil:UC-Greece},\ref{affil:IAFRT-Greece}}
	    \and Clemens Thum\inst{\ref{affil:IRAM-Granada}}
	    \and Ioannis Myserlis\inst{\ref{affil:IRAM-Granada}}
	    \and Antonio Fuentes\inst{\ref{affil:IAA-CSIC}}
	    \and Efthalia Traianou\inst{\ref{affil:IAA-CSIC}}
	    \and Jae-Young Kim\inst{\ref{affil:mpfir},\ref{affil:KNU-DAAS}}
	    \and Joana Kramer\inst{\ref{affil:mpfir}}
	    \and Rubén López-Coto\inst{\ref{affil:IAA-CSIC}}
	    \and Filippo D'Ammando\inst{\ref{affil:INAF-ira}}
	    \and M. Bernardos\inst{\ref{affil:IAA-CSIC}}
	    \and Giacomo Bonnoli\inst{\ref{affil:INAF-brera},\ref{affil:IAA-CSIC}}
	    \and Dmitriy A. Blinov\inst{\ref{affil:IAFRT-Greece},\ref{affil:UC-Greece}}
	    \and G.A. Borman\inst{\ref{affil:CAO-Russia}}
	    \and T.S. Grishina\inst{\ref{affil:SPBU}}
	    \and V.A. Hagen-Thorn\inst{\ref{affil:SPBU}}
	    \and E.N. Kopatskaya\inst{\ref{affil:SPBU}}
	    \and E.G. Larionova\inst{\ref{affil:SPBU}}
	    \and V.M. Larionov\inst{\ref{affil:SPBU}}
	    \and L.V. Larionova\inst{\ref{affil:SPBU}}
	    \and D.A. Morozova\inst{\ref{affil:SPBU}}
	    \and S.S. Savchenko\inst{\ref{affil:SPBU},\ref{affil:SAO-RAS-Russia},\ref{affil:PO-Russia}}
	    \and I.S. Troitskiy\inst{\ref{affil:SPBU}}
        \and Y. V. Troitskaya\inst{\ref{affil:SPBU}}
	    \and A.A. Vasilyev\inst{\ref{affil:SPBU}}
}
\institute{
	         Instituto de Astrofísica de Andalucía, CSIC, Glorieta de la Astronomía s/n, 18080 Granada\label{affil:IAA-CSIC}
	    \and Department of Astronomy, University of Geneva, ch. d’Ecogia 16, 1290 Versoix, Switzerland\label{affil:UniGe-astro}
	    \and Institute for Astrophysical Research, Boston University, 725 Commonwealth Avenue, Boston, MA 02215, United States\label{affil:BU-blazar}
	    \and Institute of Astrophysics, Foundation for Research and Technology - Hellas, Voutes, 70013 Heraklion, Greece\label{affil:IAFRT-Greece}
	    \and Department of Physics, University of Crete, 71003, Heraklion, Greece\label{affil:UC-Greece}
	    \and Institut de Radioastronomie Millimétrique, Avenida Divina Pastora 7, Local 20, E-18012, Granada, Spain\label{affil:IRAM-Granada}
	    \and Max-Planck-Institut für Radioastronomie, Auf dem Hügel 69, D-53121 Bonn, Germany\label{affil:mpfir}
	    \and Department of Astronomy and Atmospheric Sciences, Kyungpook National University, Daegu 702-701, Republic of Korea\label{affil:KNU-DAAS}
	    \and INAF - Istituto di Radioastronomia, Via Gobetti 101, I-40129 Bologna, Italy\label{affil:INAF-ira}
	    \and INAF Osservatorio Astronomico di Brera, Via E. Bianchi 46, 23807 Merate (LC), Italy\label{affil:INAF-brera}
	    \and Crimean Astrophysical Observatory RAS, P/O Nauchny, 298409, Russia\label{affil:CAO-Russia}
	    \and Saint Petersburg State University, 7/9 Universitetskaya nab., St. Petersburg, 199034 Russia\label{affil:SPBU}
	    \and Special Astrophysical Observatory, Russian Academy of Sciences, 369167, Nizhnii Arkhyz, Russia\label{affil:SAO-RAS-Russia}
	    \and Pulkovo Observatory, St.Petersburg, 196140, Russia\label{affil:PO-Russia}
}

\newcommand{\mykeywords}{Astroparticle physics -- Accretion, accretion disks -- Polarization -- Radiation mechanisms: general --  Galaxies: jets -- Relativistic processes}

\usepackage{fancyvrb} 

\begin{document}

        \abstract{}{}{}{}{}
\abstract
{Blazar \object{AO 0235+164}, located at redshift $z=0.94$, has undergone several sharp multi-spectral-range flaring episodes during the last decades. In particular, the episodes peaking in 2008 and 2015, that received extensive multi-wavelength coverage, exhibited interesting behavior.}
{We study the actual origin of these two observed flares by constraining the properties of the observed photo-polarimetric variability, those of the broad-band spectral energy-distribution and the observed time-evolution behavior of the source as seen by ultra-high resolution total-flux and polarimetric Very-long-baseline interferometry (VLBI) imaging.}
{The analysis of VLBI images allows us to constrain kinematic and geometrical parameters of the 7\,mm jet. We use the Discrete Correlation Function to compute the statistical correlation and the delays between emission at different spectral ranges. Multi-epoch modeling of the spectral energy distributions allows us to propose specific models of emission; in particular for the unusual spectral features observed in this source in the X-ray region of the spectrum during strong multi spectral-range flares.}
{We find that these X-ray spectral features can be explained by an emission component originating in a separate particle distribution than the one responsible for the two standard blazar bumps. This is in agreement with the results of our correlation analysis that do not find a strong correlation between the X-rays and the remaining spectral ranges. We find that both external Compton dominated and synchrotron self-Compton dominated models can explain the observed spectral energy distributions. However, synchrotron self-Compton models are strongly favored by the delays and geometrical parameters inferred from the observations.}
{}
   
        \keywords{\mykeywords}
        
        \date{Received 12 May 2023 / 
            Accepted 24 October 2023} 
        
        \maketitle

    \section{Introduction}
    
Blazars are among the most energetic objects in the universe. They are widely accepted to consist of a super massive black hole, referred to as the central engine, surrounded by an accretion disk and usually a dusty torus, and two symmetrical jets of matter emanating from the innermost vicinity of the black hole and the accretion disk.  Particles in the jet are accelerated and collimated through a variety of mechanisms, still under research, to speeds close to the speed of light. This results in highly energetic emission of radiation across the entire electromagnetic spectrum when these particles interact with the jet itself, the magnetic fields and the surrounding medium. 
For the case of blazars, the jet is pointing towards us, thus presenting relativistic effects of light aberration such us light-travel time delays, that lead to -apparent- superluminal motions, or Doppler boosting of radiation that makes them appear orders of magnitude brighter than non-blazar jets. 

    Blazars usually present a spectral energy distribution (SED) with two bumps; the first one extending from radio to optical wavelengths, or even X rays in the case of  high synchrotron-peaked (HSP) blazars; and the second one extending from X rays, or $\gamma$-rays,  to very high energy $\gamma$-rays. Synchrotron emission from the interaction of the -charged- relativistic particles of the jet with the magnetic fields in the medium is accepted to account for the first bump. 
Several scenarios exist to explain the second bump.
    In the leptonic scenario, the second bump is explained by inverse Compton effect of relativistic electrons interacting with ambient photons, and distinction is made whether these photons originate from the synchrotron emission inside the jet, in which case the mechanism is labeled as synchrotron self-Compton (SSC). On the other hand, if the photon field is originated in a region external to the jet (typically the broad line region or the dusty torus), the mechanism is labeled as external Compton (EC).
    There is an ongoing debate about the relevance of other different mechanisms, such as the so called hadronic scenarios. Frequently, combination of more than one emission mechanism is necessary to explain the observed SEDs and variability properties of the sources, even if the exact ratio of their contributions, and the origin and location of photon fields and particles involved is not sufficiently well established.
    
    The study of the variability of blazars across the spectrum, combined with the analysis of sequences of ultra-high-resolution VLBI images, has proven to be an effective way of constraining the different emission models at work in these objects (\cite{Blandford:2018}). In particular, knowledge about the exact regions around the supermassive black hole and the relativistic jet where the $\gamma$-ray emission originates is essential to discard or support different models.
    
    Regarding the location of the $\gamma$-ray emission, two main possibilities have been under discussion, differing in the distance to the central black hole (BH). The first one is the so called "close-zone" scenario, very close to the BH $(\SI{0.1}{} - \SI{1}{} \SI{}{pc})$, that was frequently used to explain the short time scales of high energy (HE) variability. However, this contradicts the coincidence of $\gamma$-ray and mm-wave outbursts that are associated to strong superluminal jet features seen in VLBI image sequences much further $\gg \SI{1}{pc}$ from the BH.  In the second one, the so called "far-zone" scenario, the emission region is located farther from the central engine, but multi-zone jet models are needed to explain the short time scales of variability reported at high- and very-high-energy $\gamma$-ray emission.

    \object{AO 0235+164}
is an extragalactic BLLac-type blazar located at redshift $z=\SI{0.94}{}$ (\cite{Cohen:1987}). It shows strong variability across all the electromagnetic spectrum and has shown interesting flaring behavior with the most recent flares occurring in 2008 and 2015 that have been studied with multi-wavelength (MWL) and Very Large Baseline Interferometry (VLBI).
    The source typically appears extremely compact at ultra-high resolution \si{7}{mm} VLBI scales (showing the whole of the emission spanning $<\SI{0.5}{mas}$) and kinematic and geometrical parameters obtained from VLBI images confirm a highly compact, narrow jet geometry pointing closely towards the observer's line of sight with a very small opening angle ($<\SI{2,4}{\degree}$) at high-speed (Doppler factor $\delta>\SI{24}{}$) which can together explain the violent outbursts reported so far (\cite{Jorstad:2001}, \cite{Weaver:2022}). \textcite{Agudo:2011} reported a detailed analysis of all measurements available up to the 2008 flare, which we extend in this paper to 2020, where we compare the two flaring episodes to shed further light about the origin and mechanisms involved in these extreme flares.
    The source has also been the subject of several previous observational campaigns, which have produced light curves showing flares in previous years, e.g. 1992 and 1998 (see \cite{Raiteri:2005}). This points to the possibility of certain level of quasi periodicity with a characteristic time scale of $\sim \SI{6}{years}$ in the behavior of the source, which can serve as a guidance while developing models of the source, even if data is not conclusive enough to settle this hypothesis and examples of non-periodic wobbling of blazar jets exist (\cite{Agudo:2012}).
 
    The 2008 flaring episode has received extensive coverage in the literature. \textcite{Agudo:2011} analyzed the flare from a multi-wavelength point of view, including polarimetric data and VLBI imaging of the source. Their results favored a SSC scenario over EC to explain the $\gamma$-ray emission and constrained the location of the emitting region at $>\SI{12}{pc}$ from the central engine.
\textcite{Ackermann:2012} also analyzed the 2008 flare, and produced a fit for the SED in the peak of the flare. In their model, EC was the dominating emission mechanism at $\gamma$-rays. However, the EC mechanism fails to explain the observed variability and the correlations between $\gamma$-ray and optical emission.
\textcite{Baring:2017} managed to reproduce the SED of \object{AO0235+164} during the peak, including the X-ray excess. 
\textcite{Wang:2020} concluded in their study that the $\gamma$-ray and mm-wave emitting zones coincided within errors and were located several parsecs from the central engine, and proposed a helical model for the jet to explain the observed polarization, without discarding other possibilities such as the shock-in-jet scenario.
     
    For this work, we have used a standard flat $\Lambda\mathrm{CDM}$ cosmological model with Hubble constant $H_0=\SI{67.66}{km / Mpc}$ as given by \cite{Planck:2018}.

    \section{Observations}
    
    We have obtained and compiled time dependent data in most available ranges of the electromagnetic spectrum, including polarimetry whenever it was possible, and VLBI polarimetric images with submilliarcsecond resolution.
    
    Our observations include 7\,mm VLBA images from the blazar monitoring program at Boston University; reduced both for total flux and polarization using AIPS (see \cite{Weaver:2022} for details about data reduction and calibration). Single dish data at \si{1}{mm} and \si{3}{mm} were obtained from the POLAMI (Polarimetric Monitoring of AGN at Millimeter Wavelengths)\footnote{{\url{https://polami.iaa.es}}} program at the IRAM 30m Telescope (\cite{polami1}, \cite{polami2}, \cite{polami3}); optical (R-band) data from Calar Alto (2.2\,m Telescope) under the MAPCAT program, the Yale University SMARTS blazar program, Maria Mitchell, Abastumani and Campo Imperatore observatories, Steward Observatory (2.3 and 1.54\,m Telescopes), the Perkins Telescope Observatory (1.8\,m Telescope), the Crimea Observatory AZT-8 (0.7\,m Telescope) and St. Petersburg State University LX-200 (0.4\,m Telescope). 
    
    Ultraviolet measurements were obtained by the \textit{Swift}-UVOT instrument. The dataset also includes X-ray data in the \SIrange{2.4}{10}{keV} range from the \textit{RXTE} satellite, and in the \SIrange{0.2}{10}{keV} energy range from \textit{Swift}-XRT; from where light curves and spectral indices were derived using a broken-power law model and the appropriate corrections for extinction. More details about the data reduction procedure from \textit{Swift} is provided in Appendix \ref{appendix:swift}. Gamma-ray data in the \SIrange{0.1}{200}{GeV} range come from the \textit{Fermi} - Large Area Telescope (LAT).

    The $\SI{\pm180}{\degree}$ polarization angle ambiguity in our R-band measurements was circumvented following the procedure described in \textcite{Blinov:2019}, which minimizes the difference between successive measurements taking also into account their uncertainty. Clusters of close observations were then shifted by an integer multiple of \SI{180}{\degree} to match the angle reported at \si{3}{mm}. This allows us for a visual comparison of the joint evolution of the optical and millimeter range polarization angles.
    
    Data from the infrared (IR) to the ultraviolet (UV) bands were corrected following the prescription by \textcite{Raiteri:2005} and the updated values by \textcite{Ackermann:2012}. This correction accounts for the local galactic extinction at $z=0$ and the intervening galaxy ELISA at $z=0.524$, as well as for ELISA's contribution to the observed emission.  When these corrections are applied, a ultraviolet bump appears in the final spectra for some epochs; as shown in \textcite{Raiteri:2005} although in disagreement with the SEDs presented in \textcite{Ackermann:2012}. It must be noted that applying different correction factors available (NED\footnote{
        The NASA/IPAC Extragalactic Database (NED)
        is operated by the Jet Propulsion Laboratory, California Institute of Technology,
        under contract with the National Aeronautics and Space Administration. \url{https://ned.ipac.caltech.edu/}.
    }, \cite{Junkkarinen:2004}, etc) also produce bumps (albeit of different intensity) but the UV bump is present in every case. Here we have followed \textcite{Raiteri:2005} when producing the final, extinction-corrected SEDs and used the updated values in \textcite{Ackermann:2012} for the extinction factors, together with the magnitudes for ELISA reported by Raiteri. A comparison with the older values by \textcite{Junkkarinen:2004} can be seen in Fig. \ref{fig:extcorr_comp}). 

    The correction of X-ray spectral data was performed using a single absorbed power law with density $N_H = \SI{2.8e21}{cm^{-2}}$ (\cite{Madejski:1996}, \cite{Ackermann:2012}), which accounts both for galactic extinction and the $z=0.524$ absorber. This value agrees with the value obtained by letting $N_H$ vary as a free parameter.

    \begin{figure*}[hptb!]
        \includegraphics[width=\textwidth]{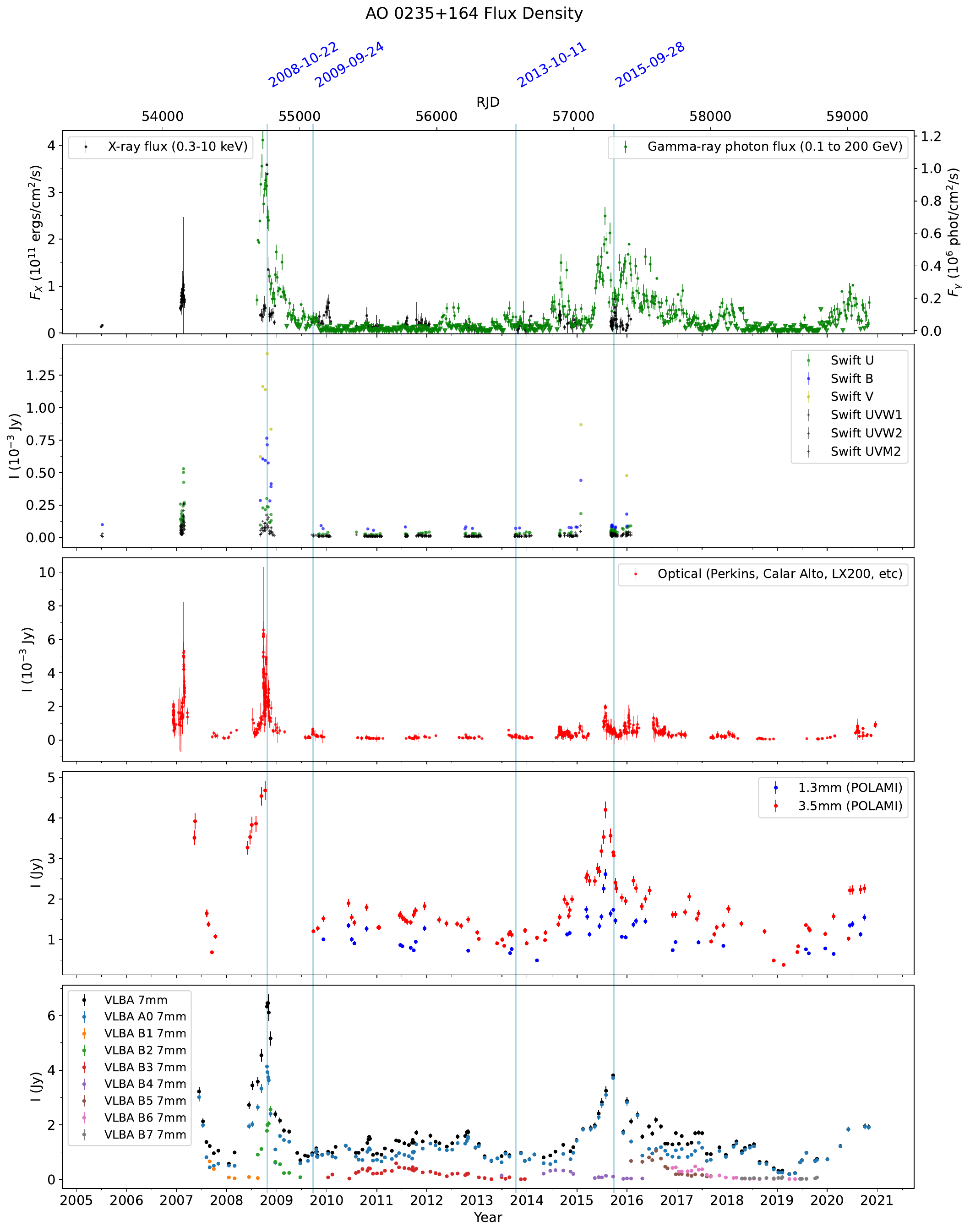}
        \caption{Light curves of \object{AO 0235+164} at different wavelengths. The vertical lines mark the epochs whose SED was analyzed in sec. \ref{sec:seds}. The top panel shows both the X-ray and $\gamma$-ray fluxes in different axes (left for X-rays and right for $\gamma$-rays, respectively).}
        \label{fig:2020_ALL_mwl_flux}
    \end{figure*}
    \begin{figure*}[hptb!]
        \includegraphics[width=\textwidth]{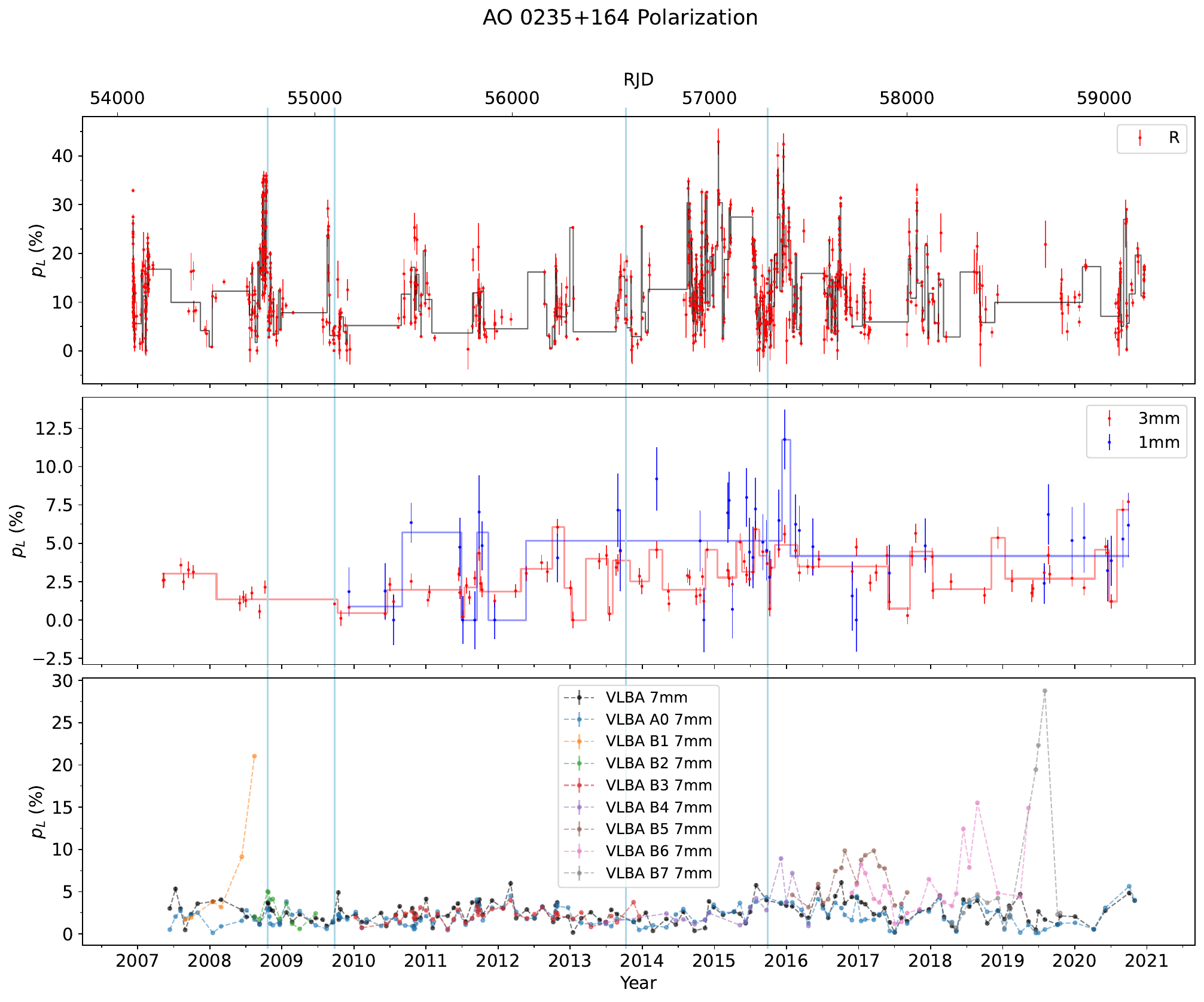}
        \caption{Polarization degree evolution of AO 0235+164 at different wavelengths. Together with experimental data, a bayesian block representation is shown superimposed for R (black line) at \SI{99.9}{\percent} confidence and for \si{1}{mm} (red) and \si{3}{mm} (blue) at \SI{90}{\percent} confidence level. The vertical lines are the same as in Fig. \ref{fig:2020_ALL_mwl_flux} and correspond to the epochs whose SED was analyzed in sec. \ref{sec:seds}.}
        \label{fig:2020_ALL_mwl_pol_I}
    \end{figure*}
    \begin{figure*}[hptb!]
        \includegraphics[width=\textwidth]{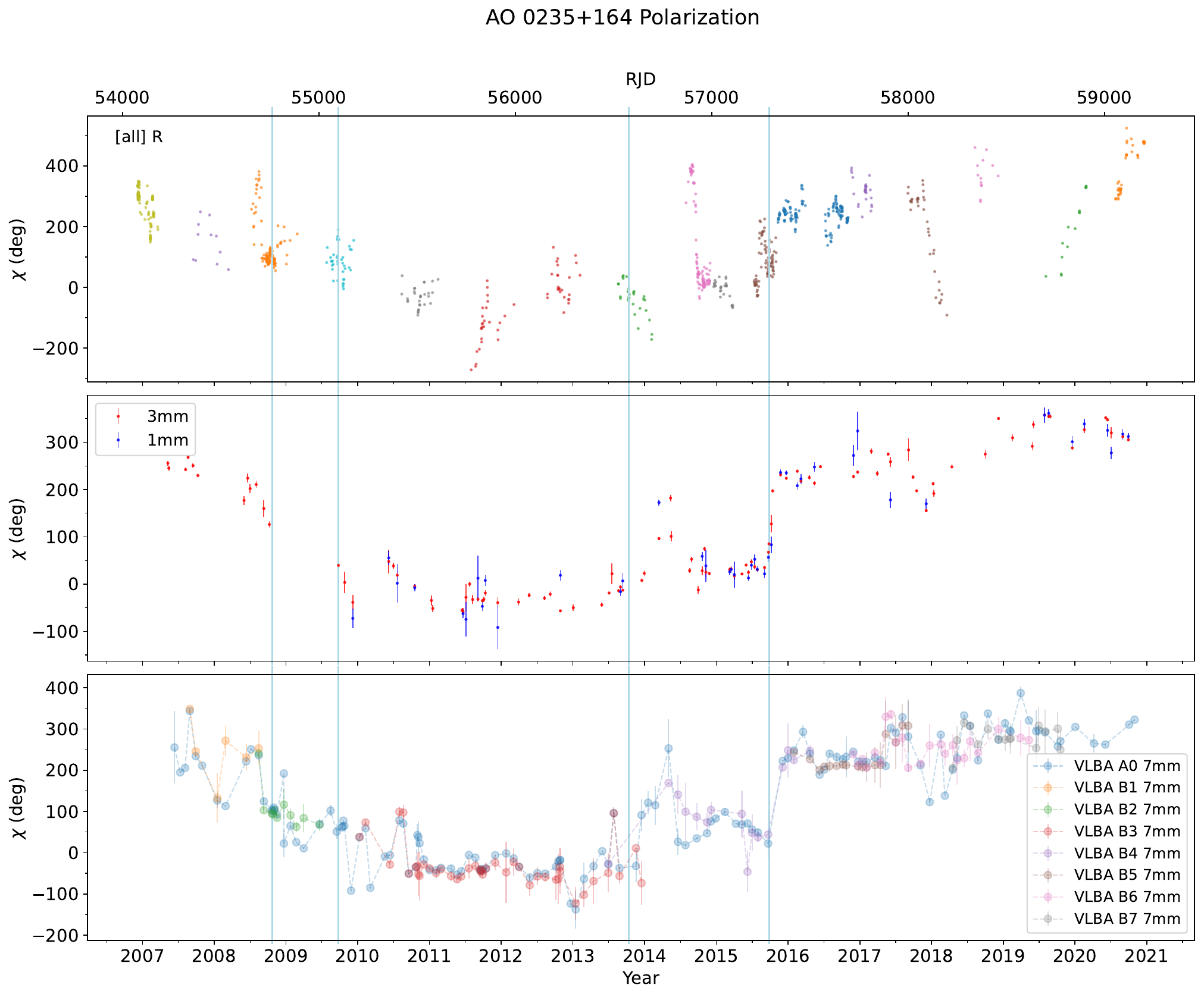}
        \caption{Polarization angle evolution of AO 0235+164 at different wavelengths. The vertical lines are the same as in Fig. \ref{fig:2020_ALL_mwl_flux} and correspond to the epochs whose SED was analyzed in sec.  \ref{sec:seds}. All points in the first box correspond to R band, the colors denote the clusters that were shifted by $n \times \ang{180}$ as mentioned in sec. \ref{sec:polarization}.}
        \label{fig:2020_ALL_mwl_pol_II}
    \end{figure*}
    
    \begin{figure*}[hptb!]
        \centering
        \includegraphics[width=0.98\textwidth]{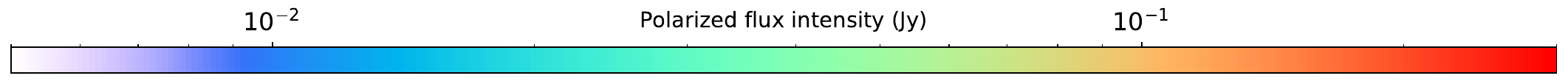}
        \includegraphics[width=0.98\textwidth]{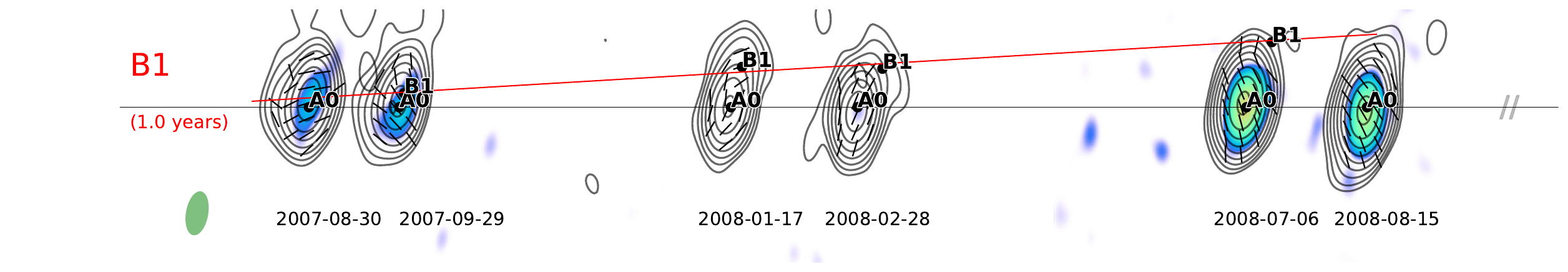}
        \includegraphics[width=0.98\textwidth]{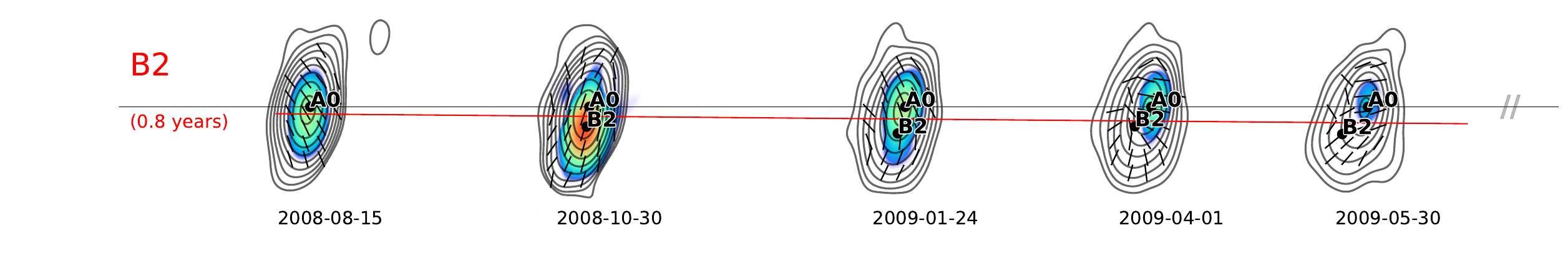}
        \includegraphics[width=0.98\textwidth]{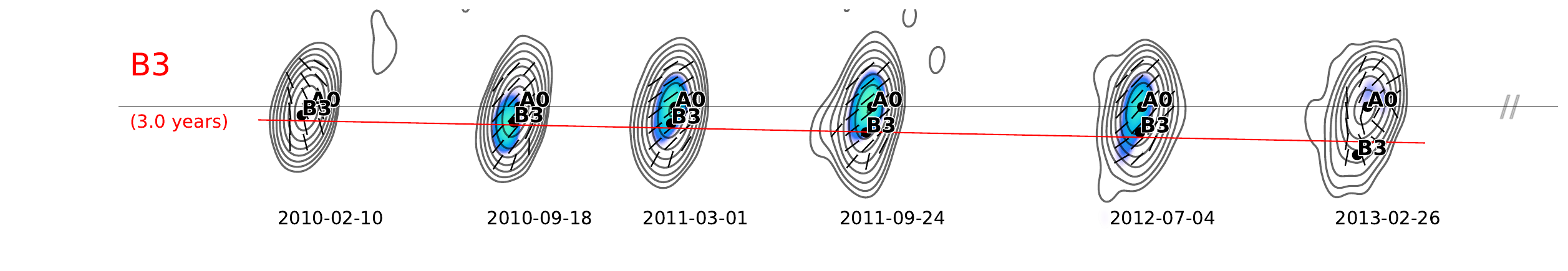}
        \includegraphics[width=0.98\textwidth]{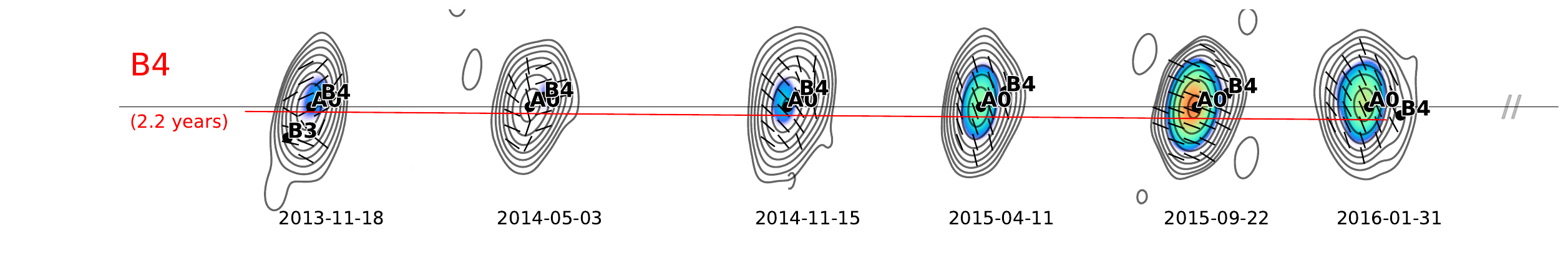}
        \includegraphics[width=0.98\textwidth]{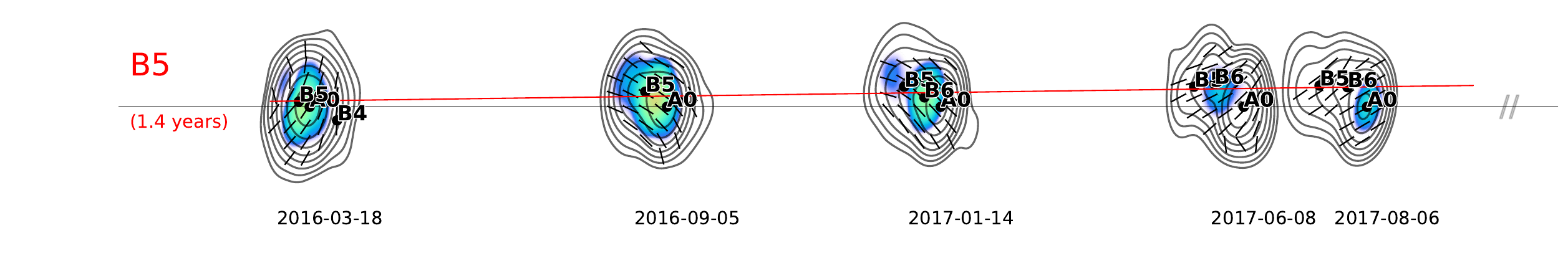}
        \includegraphics[width=0.98\textwidth]{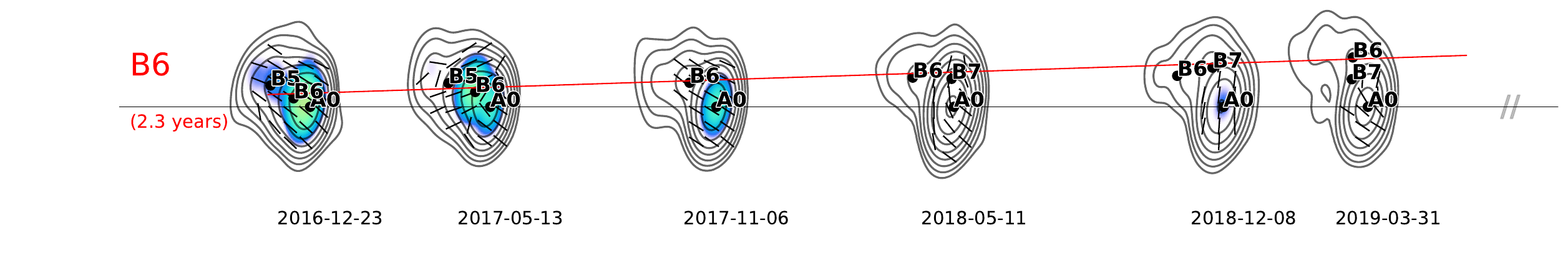}
        \includegraphics[width=0.98\textwidth]{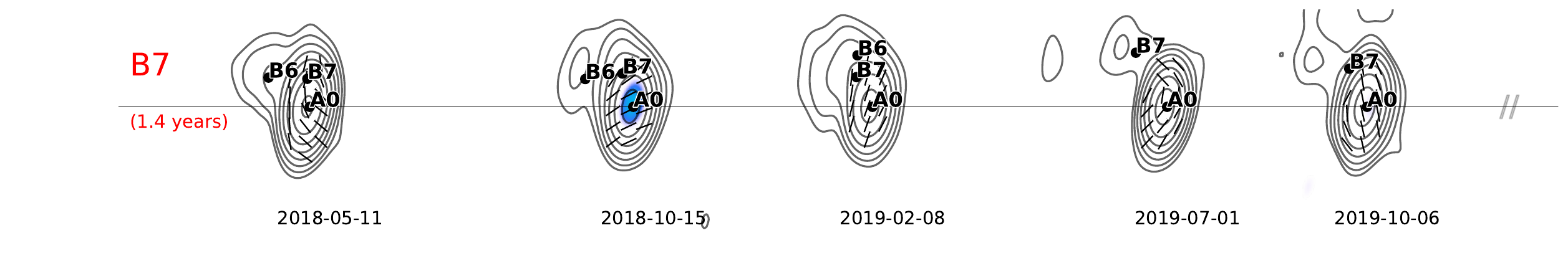}
        \caption{Selected epochs illustrating the evolution of each identified knot, showing total (contours) and polarized (color scale) intensity. The beam size is indicated as a green ellipse in the first row. Horizontal black lines indicate the position of the core A0, black line segments within the image indicate the direction of polarization (EVPA). The red line in each row is the linear fit to the knot position. It is present for every knot except B7, whose flux was too low to accurately perform a fit. For each row, the spacing between plots is proportional to time, and the total time span is different and indicated in brackets.}
        \label{fig:vlbi_knot_evolution}
    \end{figure*}

    \section{Results}
    
    \subsection{Millimeter, optical and high energies}

        The light curves at millimeter wavelengths (VLBA 7\,mm, 1\,mm, 3\,mm), optical bands (\texttt{R}, \texttt{U}, \texttt{B}, \texttt{V}), ultraviolet bands (\texttt{UVW1}, \texttt{UVW2}, \texttt{UVM2}), X-rays ($\SIrange{0.2}{10}{keV}$) and $\gamma$-rays ($ \SIrange{0.1}{200}{GeV}$) of AO 0235+164 are presented in Fig. \ref{fig:2020_ALL_mwl_flux}. Polarization degree and polarization angles at optical (R-band) and millimeter wavelengths (\si{1}{mm}, \si{3}{mm} and VLBA \si{7}{mm}) are shown in Figs. \ref{fig:2020_ALL_mwl_pol_I} and \ref{fig:2020_ALL_mwl_pol_II} respectively.
        
       Figure \ref{fig:2020_ALL_mwl_flux} show that flaring episodes happen almost simultaneously across all the electromagnetic spectrum. Variability is much more pronounced at HE, milder at optical and UV wavelengths, and softer in the millimeter and radio bands.

    \subsection{Polarization} \label{sec:polarization}

        The bayesian block representation (\cite{Scargle:2013}) of the polarization degree light curves makes it easier to discern the different behavior between quiescent and flaring states (Fig. \ref{fig:2020_ALL_mwl_pol_I}) because it represents significantly different evolution states of the source.
The source exhibits lower polarization degree at both optical and mm wavelengths during the quiescent period in between flares ($p_{L, \mathrm{R}}=\SI{9.5\pm6.0}{\percent}$, $p_{L, \mathrm{3mm}}=\SI{2.5\pm1.4}{\percent}$ from 2010 to 2014) than during flares ($p_{L, \mathrm{R}}=\SI{14.5\pm8.5}{\percent}$, $p_{L, \mathrm{3mm}}=\SI{3.34\pm1.3}{\percent}$ from 2014 to 2017).
The \si{3}{mm} polarization angles also varies more slowly during the quiescent period: from 2010 to 2014, the polarization angle at mm wavelengths remains more or less stable, while from 2014 to 2017 it performs three full $\ang{180}$ rotations, as can be seen in Fig. \ref{fig:2020_ALL_mwl_pol_II}. Rotations in the optical R-band also follow mm rotations, with a stronger variability, sometimes performing several \SI{180}{\degree} cycles while the \si{3}{mm} only varies a full cycle or a partial rotation. There is also an apparent delay of approximately a hundred days between \si{3}{mm} and R-band as can be seen in Fig. \ref{fig:2020_ALL_mwl_pol_II}.

        The direction of the EVPA (Electric Vector Position Angle) of the VLBA components (indicated in Figs. \ref{fig:vlbi_knot_evolution}, \ref{img:epoch102_2016-09-05} and \ref{img:epoch110_2017-04-16} as black lines segments overlaid with the images) coincides with the momentary direction of the jet. This alignment is in agreement with the shock-in-jet model (\cite{Marscher:2008}), where the compression of the magnetic field in the plane perpendicular to the direction of propagation, slightly off the direction of the observer, makes the electric field to align with the jet direction. This supports the association of the superluminal components ejected during flares with plane-perpendicular moving shock-waves.

    \subsection{VLBI imaging}
    
        Our study includes all available \si{7}{mm} (43GHz) VLBA total flux and polarimetric images from the Boston University Blazar Group of the sourced from 2008 to 2020 (from the VLBA-BU-BLAZAR and BEAM-ME programs\footnote{\url{https://www.bu.edu/blazars/BEAM-ME.html}}).  After reducing the data with \texttt{AIPS} \citep[see][]{Weaver:2022}, most prominent jet features were fitted to Gaussian components with \texttt{Diffmap} and then cross-identified along observing epochs. This was done for a total of 142 observing epochs. 
        
        The VLBA images show a compact, stationary component at all epochs, A0, referred here as the core. Other features can be tracked at different epochs, and their evolution traced in time. Figure \ref{fig:vlbi_knot_evolution} shows some selected epochs with the identified knot features to give a general idea of the behavior of the source in time.  Evolution curves in total and polarized flux intensity and polarization degree for the total emission and single components were later produced from the images (Figs. \ref{fig:2020_ALL_mwl_flux}, \ref{fig:2020_ALL_mwl_pol_I} and \ref{fig:2020_ALL_mwl_pol_II}) using the aforementioned identification.
        
        The flux evolution shown in Fig. \ref{fig:2020_ALL_mwl_flux} at all wavelengths, also containing the light curves from the integrated VLBA \si{7}{mm} maps, allows us to distinguish two clear \textit{flaring} periods, whose peaks of activity occurred in October 2008 and July 2015 respectively. The 2008 flare is associated with the B2 jet feature that developed southwest of the core (A0). In contrast, the 2015 flare is associated with jet components B5 and B6 that developed northwest. 
Other weaker components not associated with the main outbursts (e.g. B4), also propagate in different directions.
This hints at a possible rotation or wobbling of the jet and supports a helical jet model, and might be associated to a pseudo periodic behavior as proposed by \textcite{Raiteri:2005}. 
All VLBI jet components have lifetimes lasting several years. During their lifetimes, we observe them propagating quasi-ballistically in the same direction relative to the core, with trailing components maintaining the same direction of propagation as their leading component as well as for its EVPA alignment (parallel to the direction of propagation in the plane of the sky). It is therefore clear that jet nozzle changes direction with time, as in each flaring episode the direction of propagation of the associated superluminal components is radically different for every on of these episodes.

    \subsection{Differences and similarities between 2008 and 2015 flares}

        The flare in 2008, reported by \textcite{Agudo:2011}, featured a superluminal component ejected from the core during June 2008 which kept separating from the core in the southeast direction during the following months, until the last months of 2009, when the component went practically extinct. \citeauthor{Agudo:2011} also reported correlations between all wavelengths, from radio to $\gamma$ rays. The flare in 2008 reached a maximum in October 2008, peaking at \SI{4.7}{Jy} in \si{3}{mm} and a magnitude of \num{14.2}, having increased its brightness by more than a factor of $60$ in optical R band with respect to its quiescent state. Although a general correlation was reported between optical and $\gamma$-ray bands, the correlation was poorer and less detailed during the main burst.

        \begin{figure}[hptb!]
            \centering
            \includegraphics[width=0.9\linewidth]{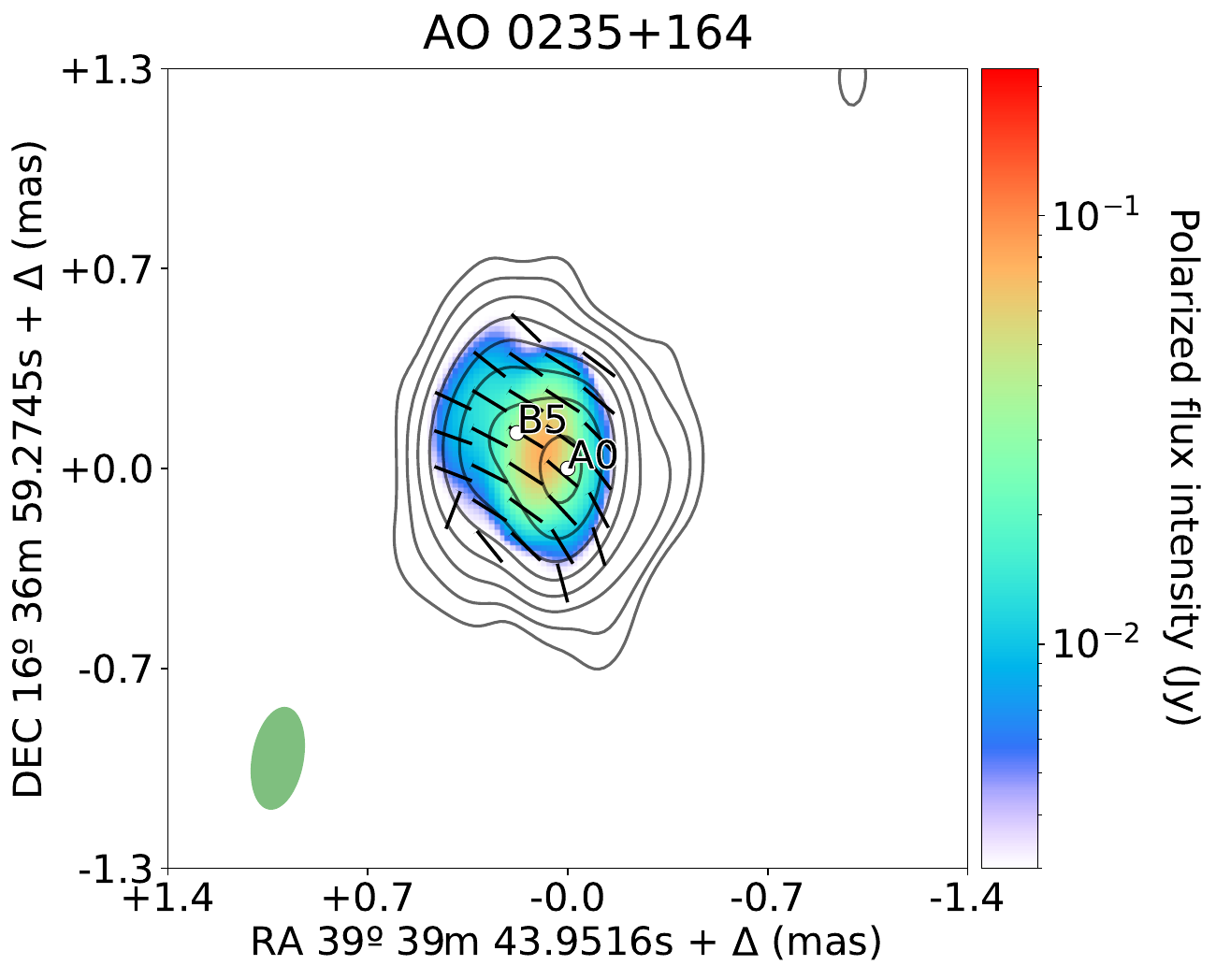}
            \caption{Epoch 102 (2016-09-05) VLBI \si{7}{mm} image, showing total flux intensity (contours) and polarized flux intensity (color scale). Black line segments overlaid in the image represent the Electric Vector Position Angle (EVPA). The green ellipse in the lower left corner represents the beam size. The image showcases component B5 close to the peak of the 2015 flaring episode, and demonstrates how the polarization angle is aligned with the direction of propagation.}
            \label{img:epoch102_2016-09-05}
        \end{figure}
    
        \begin{figure}[hptb!]
            \centering
            \includegraphics[width=0.9\linewidth]{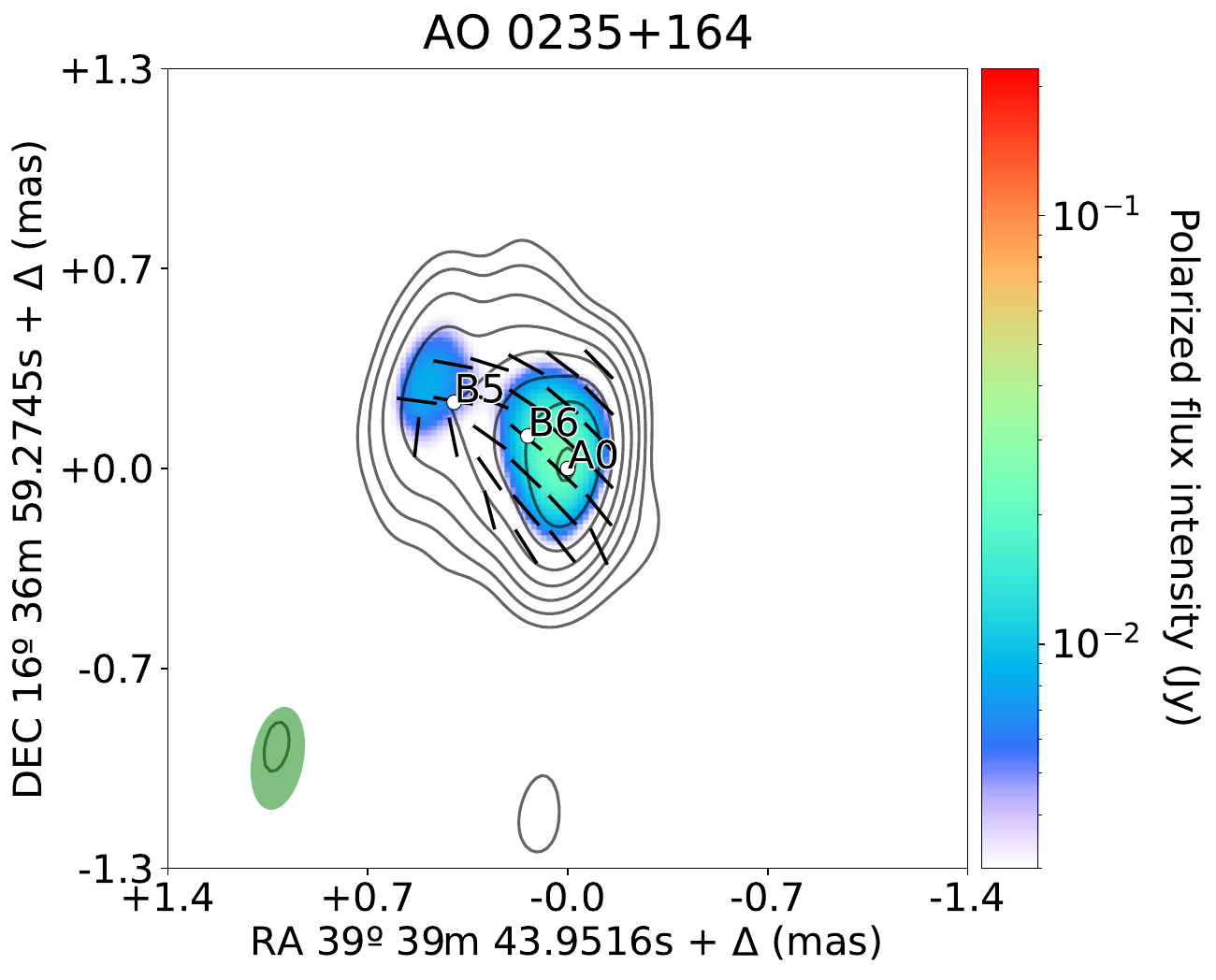}
            \caption{Epoch 110 (2017-04-16) VLBI \si{7}{mm} image. The image shows the trailing component B6 moving in the same direction of B5, maintaining the alignment of the polarization angle (black line segments) with the momentary direction of the jet (northeast).}
            \label{img:epoch110_2017-04-16}
        \end{figure}
        
        The second flaring activity begins around Fall 2014, with a brightening of the core visible at \si{7}{mm}. The flux densities at all wavelength peak a year later, around Fall 2015, at all wavelengths. Comparatively, this flare is dimmer than the previous one of 2008, reaching \SI{4.2}{Jy} in \si{3}{mm} and 1.3 magnitudes less in optical R band.  A plausible explanation for this will be given below through the interpretation of the flare in terms of emission zones and shocks.
        
        VLBI images reveal that the component responsible for the brightening of the core during the 2015 flare, named B5, originates in the core less than a few weeks before the brightening in total flux density begins. After it separates from the core, the component travels outwards at a position angle of $\sim \ang{45}$ while increasing its brightness, and peaks around Fall 2016. Figure \ref{img:epoch102_2016-09-05} shows the total and polarized intensity image of the source during nearest to this event. This is around a year later than the total flux peak.
        The interpretation for this is that the initial brightening of the total flux is due to the interaction of the component with the core, while the ensuing brightening of the component is due to acceleration or a change of viewing angle.
                
        Some months later another, a weaker component, named B6, originates from the core. This component follows the path of B5, peaking around April 2017 (Fig. \ref{img:epoch110_2017-04-16}), while B5 is still visible. This component is the one responsible for the subflare of 2017, visible at all wavelengths in total flux. The behavior of this component is compatible with that of a trailing component of B5 as described by \textcite{Agudo:2001}. By 2019 all activity had ended, with a minimum reached around March 2019.

\begin{figure} \centering
            \includegraphics[width=1.0\linewidth]{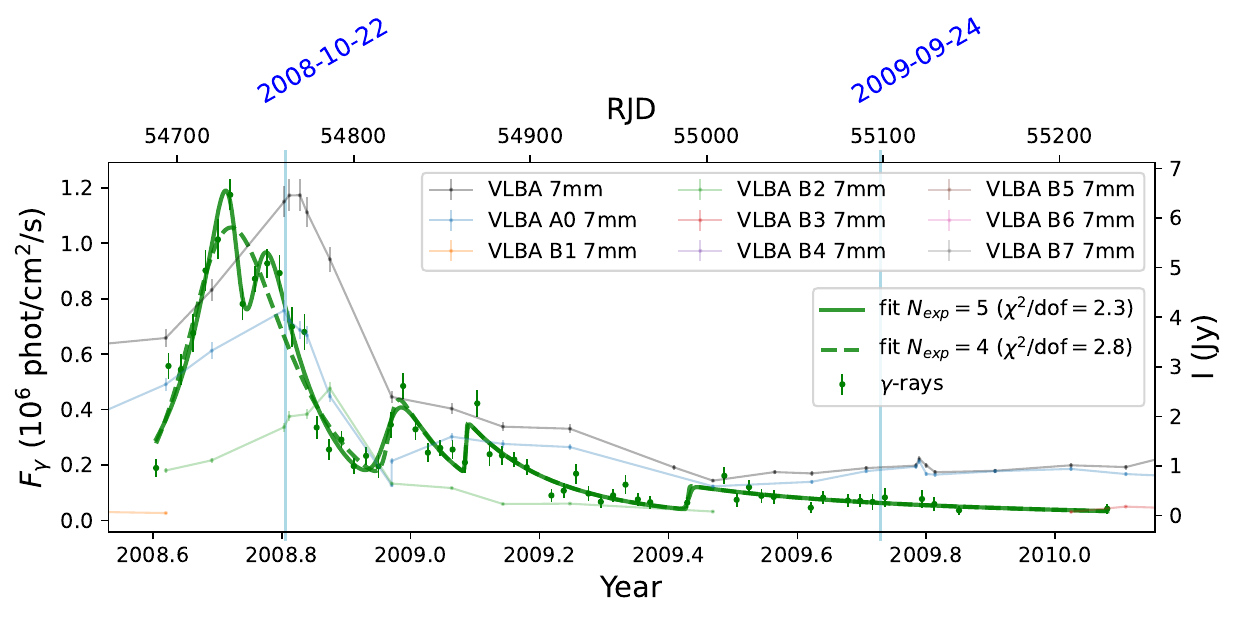}
            \includegraphics[width=1.0\linewidth]{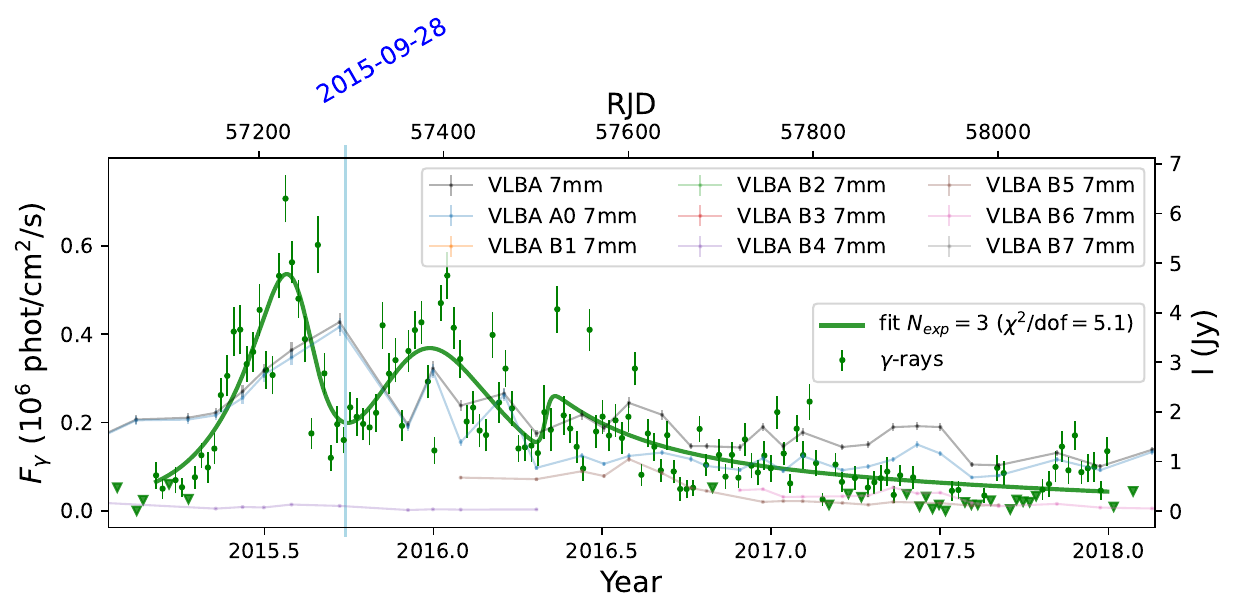}
            \includegraphics[width=0.9\linewidth]{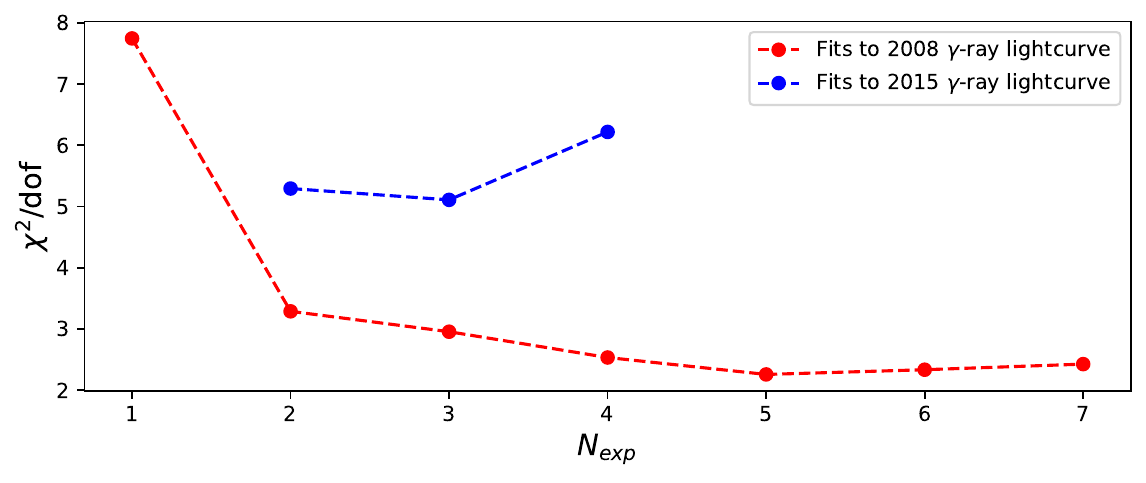}
            \caption{Fits to the profiles of the 2008 (top) and the 2005 (middle) $\gamma$-ray flares by exponential functions as described by eq. \ref{eq:smooth_exps}. The selected number of terms $N_{exp}$ were the ones such that the reduced $\chi^2$-statistic was minimized (bottom).  The best-fit values are given in Tables \ref{tb:timeprofile_2008_fit_results} and \ref{tb:timeprofile_2015_fit_results}. An alternative fit is given for the 2008 flare with a similar $\chi^2$, that accounts for the double peak at the beginning with a single exponential term. The vertical lines mark the epochs whose SED was analyzed in sec. \ref{sec:seds}, as in Fig.  \ref{fig:2020_ALL_mwl_flux}.}
            \label{fig:flares_2008_2015_selected_fits}
    \end{figure}

        The two flares present different time profiles at $\gamma$-ray energies. A comparison of these profiles can be found in 
Fig. \ref{fig:flares_2008_2015_selected_fits}, 
        where also the 7mm flux of each identified  VLBI component  is shown. We have modeled the $\gamma$-ray profiles by fitting to standard exponential shapes given by (\cite{Abdo:2010})
\begin{equation}
            y(t)_{N_{exp}} = \sum_i^{N_{exp}} A_i \cdot { \left(  e^{(t-{t_c}_i)/{t_r}_{i}} + e^{(t-{t_c}_i)/{t_d}_{i}} \right) }^{-1} 				
            ~,
            \label{eq:smooth_exps}
        \end{equation}
which allowed us to derive the rising and decaying times of each subflare,  ${t_r}_{i}$ and ${t_d}_{i}$. Then, their asymmetry factor, defined as 
       \begin{equation}
           \xi_i = \dfrac{{t_r}_i-{t_d}_i}{{t_r}_i+{t_d}_i}	
           ~,
           \label{eq:flare_sym_factor}
       \end{equation}
could be computed. An asymmetry factor close to zero corresponds to the case of a perfectly symmetric flare. There exists some uncertainty in the number of exponential terms to use, since the source shows strong variability in timescales shorter than our binning allows to track. The large binning used, of 7 days, was necessary to accommodate periods of low flux. Still, it can be seen that the source displays significant variations of flux even in these intervals. 
The value of $N_{exp}$ chosen was the one that minimized the reduced $\chi^2$-statistic. The results of both fits are shown in Fig. \ref{fig:flares_2008_2015_selected_fits}, and the corresponding parameters in Tables \ref{tb:timeprofile_2008_fit_results} and \ref{tb:timeprofile_2015_fit_results}.
The distance from the fitted $\gamma$-ray subflare maximum to the 7mm maximum in 2015 is \SI{52}{days} ($\pm\SI{8}{days}$), a similar delay to the one found in the DCF analysis in sec. \ref{sec:correlations} ($\tau_{R, \gamma} = \SI{2}{days}$, $\tau_{R, \mathrm{7mm}} = \SI{64\pm 4}{days}$). 
        
        \begin{table*}[htbp!]
            \centering
\caption{Parameters for the fit to the $\gamma$-ray lightcurve of the 2008 flaring episode to functions of shape given by eq. \eqref{eq:smooth_exps}. The resulting reduced $\chi^2$-statistic for the fit is shown, and also the computed symmetry factor $\xi_i$ for each subflare. Some values could not be computed. Upper limits are indicated with '$<$'. The result can be seen in Fig. \ref{fig:flares_2008_2015_selected_fits}.}
            \label{tb:timeprofile_2008_fit_results}
\begin{tabular}{rrrrrrr}
    \toprule
    {$N_{exp}$} & 
    ${A}_i$ ($\times 10^{6}$) & 
    ${t_r}_i$ [\unit{days}] & 
    ${t_d}_i$ [\unit{days}] & 
    ${t_c}_i$ [\unit{year}] & 
    ${\chi^2} / \mathrm{d.o.f}$ &    $\xi$\\
    \midrule
    4 & 
        \makecell[rt]{\num{2.0} (\num{0.1}) \\ \num{0.4} (\num{0.1}) \\ \num{0.2} ({$...$}) \\ \num{0.09} ($...$)} & 
        \makecell[rt]{\num{17.8} (\num{4.3}) \\ $<\num{2.6}$ \\ $<\num{26}$ \\ $<\num{200}$} & 
        \makecell[rt]{\num{36.4} (\num{3.2}) \\ \num{37.8} (\num{20}) \\ $...$ \\ $...$} & 
        \makecell[rt]{\num{2008.698} (\num{0.014}) \\ \num{2008.968} (\num{0.006}) \\ \num{2009.085} ({$...$}) \\ \num{2009.430} ($...$)} & 
        2.8 & 
        \makecell[rt]{\num{-0.3} \\ \num{-0.9} \\ \num{-1.0} \\ \num{-1.0}} \\
    \midrule
    5 & 
        \makecell[rt]{\num{1.7} (\num{0.4}) \\ \num{1.5} (\num{0.4}) \\ \num{0.5} (\num{0.1}) \\ \num{0.2} (\num{4e+03}) \\ \num{0.09} (\num{4e+01})} &
         \makecell[rt]{\num{25.7} (\num{2.1}) \\ \num{5.3} (\num{5.0}) \\ \num{4.6} (\num{2.9}) \\ $<\num{17.2}$ \\ $<\num{290}$} & 
         \makecell[rt]{$<\num{4.4}$ \\ \num{27.9} (\num{4.7}) \\ \num{37.7} (\num{21}) \\ $...$ \\ $...$} & 
         \makecell[rt]{\num{2008.728} (\num{0.008}) \\ \num{2008.758} (\num{0.007}) \\ \num{2008.965} (\num{0.012}) \\ \num{2009.086} ($...$) \\ \num{2009.432} (\num{44})} &
          2.3 & 
          \makecell[rt]{\num{0.7} \\ \num{-0.7} \\ \num{-0.8} \\ \num{-1.0} \\ \num{-1.0}} \\
    \bottomrule
\end{tabular}
        \end{table*}
       \begin{table*}[htbp!]
           \centering
\caption{Parameters for the fit to the $\gamma$-ray lightcurve of the 2015 flaring episode. The resulting reduced $\chi^2$-statistic for the fit is shown, and also the computed symmetry factor $\xi_i$ for each subflare. The result can be seen in Fig. \ref{fig:flares_2008_2015_selected_fits}.}
           \label{tb:timeprofile_2015_fit_results}
\begin{tabular}{rrrrrrr}
\toprule
{$N_{exp}$} & 
               ${A}_i$ ($\times 10^{6}$) & 
               ${t_r}_i$ [\unit{days}] & 
               ${t_d}_i$ [\unit{days}] & 
               ${t_c}_i$ [\unit{year}] & 
               ${\chi^2} / \mathrm{d.o.f}$ &    $\xi$\\
\midrule
3  & \makecell[rt]{\num{0.9} (\num{0.2}) \\ \num{0.7} (\num{0.1}) \\ \num{0.15} (\num{0.07})} & \makecell[rt]{ \num{58} (\num{12}) \\ \num{49} (\num{35}) \\ \num{3.2} (\num{6.4})} & \makecell[rt]{ \num{18.2} (\num{7.7}) \\ \num{86} (\num{48}) \\ \num{490} (\num{340})} & \makecell[rt]{\num{2015.605} (\num{0.022}) \\ \num{2015.941} (\num{0.122}) \\ \num{2016.332} (\num{0.018})}  & 5.1 & \makecell[rt]{\num{0.5} \\ \num{-0.3} \\ \num{-1.0}} \\
\bottomrule
\end{tabular}
        \end{table*}

        During the 2015 flaring episode, the secondary flares in $\gamma$-rays are contemporaneous with the appearance of 7mm VLBI components. In particular, the first, second and third maxima happen at approximately the same time A0 rebrightens and the B5 and B6 components appear. 
This suggests that these emissions are spatially related.The failure in finding  components responsible for the subsequent fitted subflares might be due to the difficulty in fitting  low-flux VLBI components, to the uncertainty in the $\gamma$-ray lightcurve, or a combination of both. 

        The same can be said about the first and second subflares of the 2008 flaring episode, where it seems that the brightening of the core A0 and the appearance of the B2 component are related to the first and second maximums at $\gamma$-rays, taking into account the aforementioned delay. 
        
        The $\gamma$-ray subflare that can be seen in Fig. \ref{fig:2020_ALL_corrs_flux} one year before the start of the 2014 flaring episode might be related to the B4 component in the same way, but this relation and the associated delay is less clear. This could potentially be the case also with the B1 component and the flare that can be seen in 2006, before the 2008 episode, in all wavelengths except $\gamma$-ray (due to the lack of observational data).
Altogether, it seems that there is a direct relationship between the appearance of the 7mm VLBI components and the successive $\gamma$-ray subflares.
        
        For the 2015 flaring episode, the rising and decaying times shown in Table \ref{tb:timeprofile_2015_fit_results}, as low as  $\sim\SI{18}{days}$ (taking into account only the two strongest subflares) are compatible with the sizes found in the SED modeling of sec. \ref{sec:seds} (Table \ref{tb:sed_models_ec_plc}), which limits to $\SI{15}{days}$ the shortest timescale where significant variations of flux can occur (eq. \ref{eq:Rb_from_delta}). These other values were obtained only from the modeling of the SEDs and the two analysis are completely independent. In the case of the 2008 flaring episode, however, the shorter times ($\sim \SI{4}{days}$) are in tension with the region sizes. This might be explained by unaccounted sources of $\gamma$-ray variability originating in smaller regions. However, it could also be caused by a wrong estimation of the rising and decaying times. It is possible to produce a fit with a single term accounting for the initial double peak  that has a similar $\chi^2$-square statistic, but rising and decaying times of \num{18} and \num{36} \si{days} (dashed line in Fig. \ref{fig:flares_2008_2015_selected_fits}). 
In any of the ways, the observed delays between $\gamma$ and mm might be explained by a combination of adiabatic expansion and cooling time (\cite{Tramacere:2022}).
        
    \section{Analysis}

    \subsection{Kinematic parameters of the VLBI jet components}
    
        From the VLBI imaging data, some kinematics parameters associated to the different visible emission zones were computed following the procedure described in \textcite{Weaver:2022}. These include $t_0$, the ejection time, which is the time where the extrapolated trajectory of the component crosses the core; $t_{var}$, the timescale of variability, which is the timescale of the dimming of the component; $\beta_{app}$, the apparent speed in units of \SI{}{c}; $\delta_{var}$, the variability Doppler factor; $\Gamma$, the Lorentz factor; and $\Theta$, the viewing angle of the jet component. 

        \begin{figure}[htbp!]
            \centering
            \includegraphics[width=0.99\linewidth]{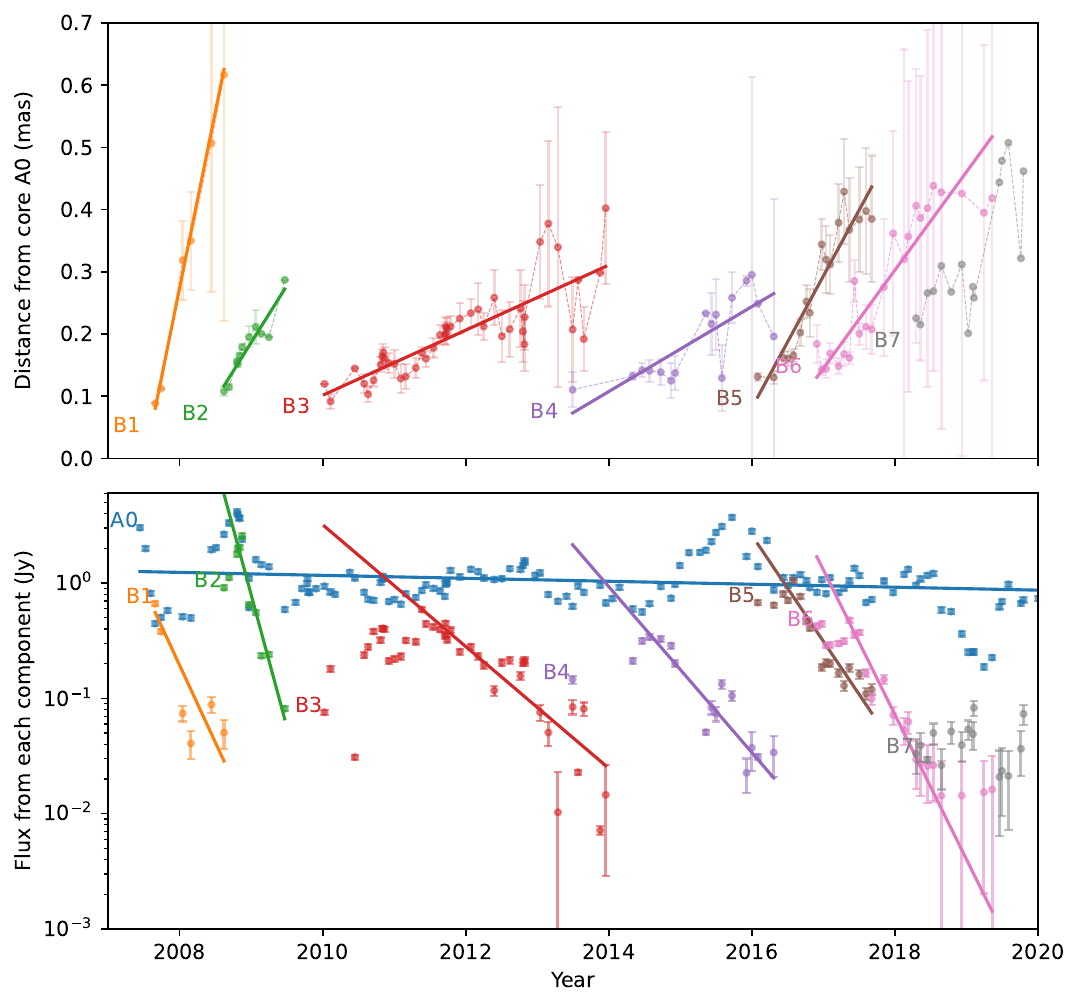}
            \caption{Observed distance from core (up) and flux density (down) for every one of the identified component as a function of time, together with a linear weighted fit to knot distance and logarithmic fit to flux. The fit to the flux is done taking into account only the points after the peak of emission of each component. This was done for all components except for B7, which due to its low flux, did not have enough points after the peak with low enough uncertainty to perform a fit.}
            \label{fig:knot_dist_n_flux_fits}
        \end{figure}

        The identified knot features in every epoch were traced and their positions adjusted to a linear fit, from which their speed were obtained, and also their flux was fitted to a decaying exponential function $F=F_{0}\exp{(-t/t_{var})}$, obtaining their timescale of variability (Fig. \ref{fig:knot_dist_n_flux_fits}).

        The Doppler factor and apparent speed were then computed as (\cite{Jorstad:2005}, \cite{Casadio:2015})
        \begin{align}
           \delta_{var}  &=  \dfrac{ 1.6 \,  a_{S_{max}} \, d_L }{c t_{var} (1+z)}	 \\
            \beta_{app}  &=  \dfrac{v_r  d_L}{c(1+z)}   			
        \end{align}
        where $a_{S_{max}}$ is the FWHM of the component measured at its maximum flux, $v_r$ is the radial velocity of the knot, and $d_L$ is the luminosity distance, but following the more robust approach found in \textcite{Weaver:2022} and using the value of $t_{var}$ obtained from the fit. From these, the Lorentz bulk factor,
        \begin{equation}
            \Gamma =  \dfrac{1}{2\delta_{var}} \left(\beta_{app}^2+\delta_{var}^2+1\right)
        \end{equation}
        and the viewing angle,
        \begin{equation}
            \tan{\Theta} = \dfrac{2 \beta_{app}}{\beta_{app}^2+\delta_{var}^2-1}	
        \end{equation}
        could be computed.
        
        Our results for these parameters (Table \ref{tb:kinematics}) agree with those of \textcite{Weaver:2022} within the expected margin of error associated with the identification of the components in the VLBA images.
        
        \begin{table*}[htbp!]
\centering
\caption{Kinematics parameters for identified knots of 0235+164 ($[v_r] = \SI{}{mas/year}$, $\langle a \rangle = \SI{}{mas}$, $t_{var} = \SI{}{year}$, $[\Theta] = \SI{}{\degree}$, other units are dimensionless.)}
\label{tb:kinematics}

\nottoggle{refereever}{
    \begin{adjustbox}{center, width=0.9\textwidth, margin=0 0 0 0}
}{
    \begin{adjustbox}{center, width=0.6\textwidth, margin=0 0 0 0}
}

   \begin{tabular}{lrrrrrrrrrrrrrrrr}
       \toprule
       {Component} &  $v_r$ &  $\sigma_{v_r}$ &   $t_0$ &  $\sigma_{t_0}$ &  $\langle a \rangle$ &  $\sigma_{a}$ &  $t_{var}$ &  $\sigma_{t_{var}}$ &  $\delta_{var}$ &  $\sigma_{\delta_{var}}$ &  $\beta_{app}$ &  $\sigma_{\beta_{app}}$ &  $\Gamma$ &  $\sigma_{\Gamma}$ &  $\Theta$ &  $\sigma_{\Theta}$ \\
       \midrule
       B1 &   0.57 &            0.05 & 2007.52 &            0.09 &                 0.29 &          0.02 &       0.32 &                0.02 &           70.5 &                     5.4 &          28.28 &                    2.55 &     40.9 &               6.3 &      0.56 &               0.06 \\
       B2 &   0.18 &            0.02 & 2007.98 &            0.12 &                 0.16 &          0.01 &       0.19 &                0.00 &           67.8 &                     3.6 &           9.14 &                    1.09 &     34.5 &               4.0 &      0.22 &               0.03 \\
       B3 &   0.05 &            0.01 & 2008.06 &            0.11 &                 0.17 &          0.01 &       0.82 &                0.01 &           16.8 &                     1.0 &           2.61 &                    0.29 &      8.6 &               1.1  &      1.04 &               0.12 \\
       B4 &   0.07 &            0.01 & 2012.42 &            0.21 &                 0.15 &          0.01 &       0.60 &                0.02 &           20.3 &                     1.7 &           3.40 &                    0.73 &     10.5 &               1.9 &      0.92 &               0.20 \\
       B5 &   0.21 &            0.02 & 2015.61 &            0.09 &                 0.24 &          0.01 &       0.47 &                0.01 &           39.8 &                     2.2 &          10.57 &                    0.94 &     21.3 &               2.5 &      0.71 &               0.07 \\
       B6 &   0.16 &            0.04 & 2016.08 &            0.22 &                 0.28 &          0.01 &       0.35 &                0.02 &           63.5 &                     5.4 &           7.87 &                    1.76 &     32.3 &               6.0 &      0.22 &               0.05 \\
       \bottomrule
   \end{tabular}
   
\end{adjustbox}         \end{table*}
    
        The results agree with the observed behavior of the flares. The estimated viewing angle for the component responsible for the 2008 flare (B2) is \SI{0.2}{\degree}, between three and four times smaller than the \SI{0.7}{\degree} of the component responsible for the 2015 flare (i.e. B5). This consistently explains the lower brightness observed in 2015 as being caused by a weaker Doppler boosting of the emission. The viewing angle for the secondary component B6 is similar to the one of B2; but it's apparent speed is much less than any of the others.

    \subsection{Correlations across the spectrum} \label{sec:correlations}
    
       Correlations between the different lightcurves were computed using \texttt{MUTIS}\footnote{MUltiwavelength TIme Series.  A Python package for the analysis of correlations of light curves and their statistical significance. \url{https://github.com/IAA-CSIC/MUTIS}}. In particular, since we are dealing with irregularly sampled signals (light curves), we compute the Discrete Correlation Function (DCF) as proposed by \textcite{Welsh:1999}, which is a normalized and binned DCF. 
           
        A uniform bin size of $\SI{20}{days}$ was used for all correlations. The choice of a uniform bin size was done so that the results of different correlations could be easily compared, the specific value of \SI{20}{days} was done so that it was large enough to have to have statistics in any bin but short enough that the the correlations were not smoothed too much and peak positions could still be determined. To confirm the robustness of our choice, we have also reproduced our analysis with bin-sizes from \SI{10}{} to \SI{30}{days}, obtaining similar results (except for some bins disappearing due to not having enough points to compute the correlation, as we discarded bins where the number of pairs was less than 11).
        
        To estimate the significance of the correlations, a Monte-Carlo approach was used, generating $N=2000$ synthetic light curves for each signal. Randomization of the Fourier transform was used for mm-wavelengths, while for optical and $\gamma$-ray data we modeled the signals as Orstein-Uhlenbeck stochastic processes (\cite{Bonnoli:2004}). The uncertainties of the correlations were estimated using the uncertainties of the signals again with a Monte-Carlo approach. 
        
    \subsubsection*{Correlation of the whole period (2007 to 2020)}
    
\begin{figure*}\centering
            \includegraphics[width=\textwidth]{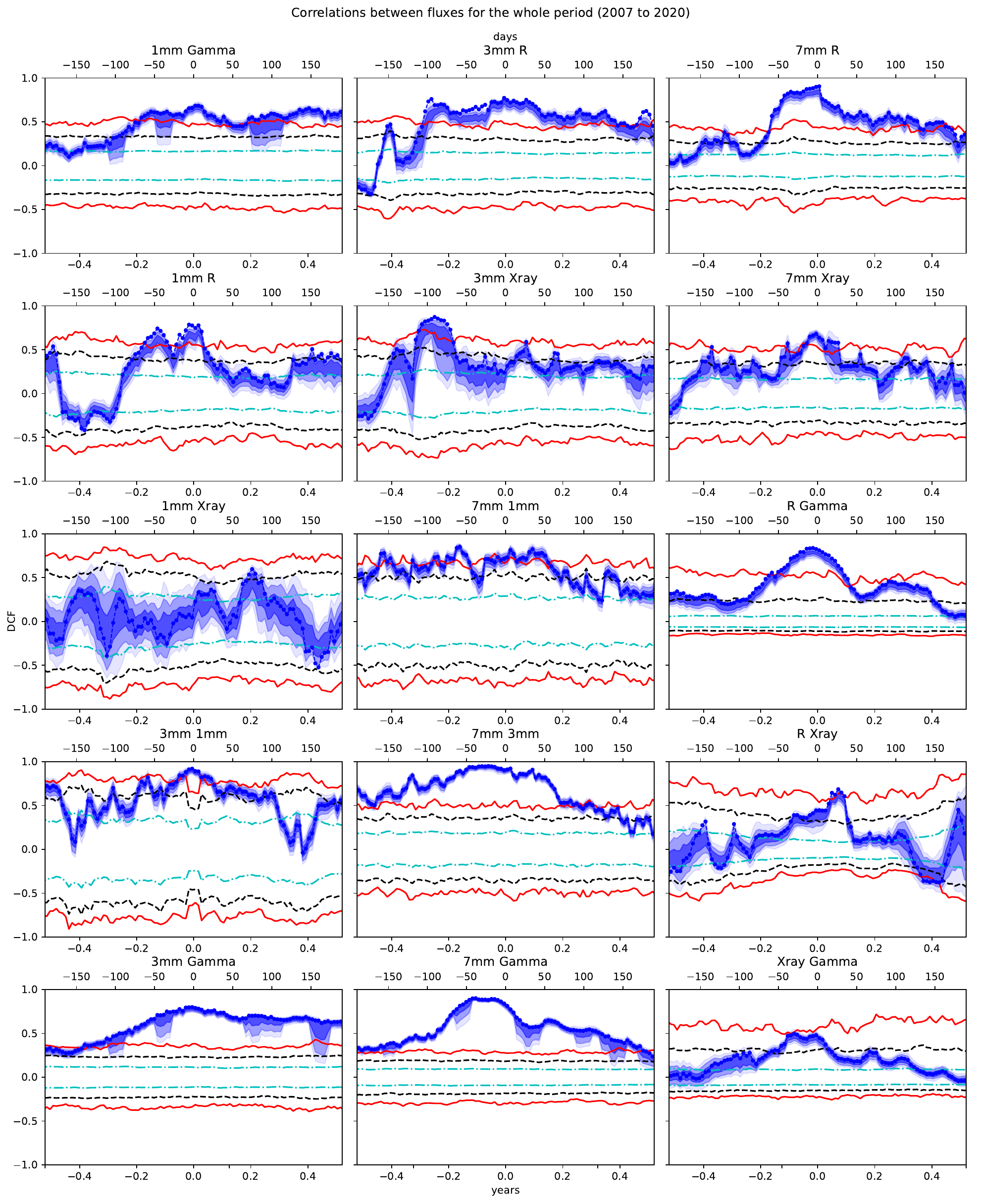}
            \caption{Correlations between fluxes across all wavelengths. Horizontal lines show significance levels for $1\sigma$, $2\sigma$, and $3\sigma$, computed using $N=2000$ synthetic light curves as described in Sec. \ref{sec:correlations}. The DCFs here are computed using the whole period of available data, from 2007 to 2020.}
            \label{fig:2020_ALL_corrs_flux}
        \end{figure*}

        The results from DCFs of the whole available period of data (2007 to 2020) show a clear correlation between flux at almost all wavelengths ($>3\sigma$), with most peak positions close to zero (Fig. \ref{fig:2020_ALL_corrs_flux}). The X-ray band is an exception to this, showing no significant ($>3\sigma$) correlation close to zero with some of the other bands. More hints about this will be provided in the following.
            
\begin{figure*}\centering
            \includegraphics[width=\textwidth]{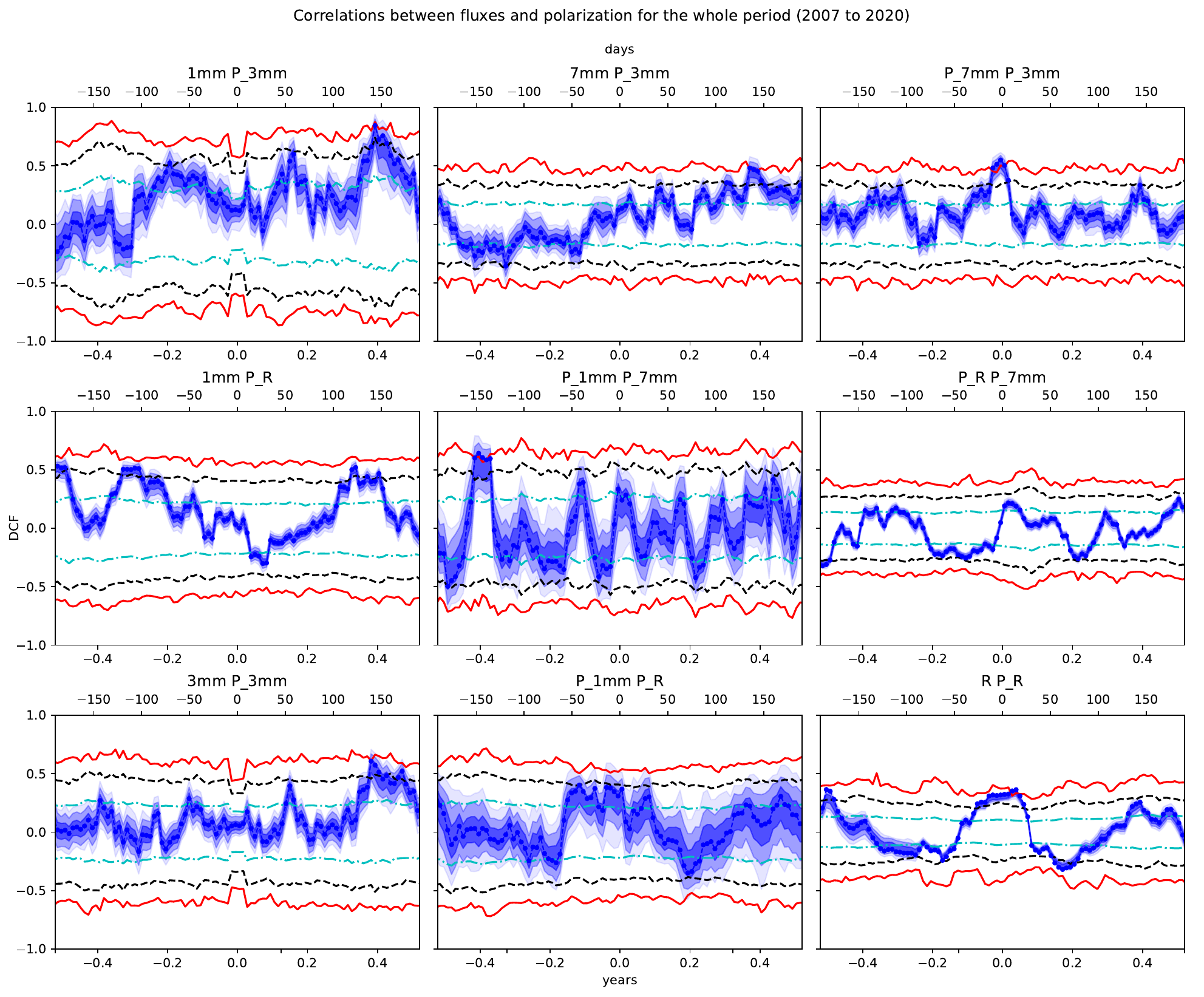}
            \caption{Correlations between fluxes and polarization degree across all wavelengths. Horizontal lines show significance levels for $1\sigma$, $2\sigma$, and $3\sigma$, computed using $N=2000$ synthetic light curves as described in Sec. \ref{sec:correlations}. The DCFs here are computed using the whole period of available data, from 2007 to 2020.}
            \label{fig:2020_ALL_corrs_pol}
        \end{figure*}

        The correlation between the polarization degree and total flux is clear for the R band, where it shows a statistically significant maximum near zero, but is not certain in the  other bands, possibly explained by the sparser sampling and larger errors (Fig. \ref{fig:2020_ALL_corrs_pol}).
        
    \subsubsection*{Correlations of flaring episode (2014 and 2017)}
        
\begin{figure*}\centering
            \includegraphics[width=\textwidth]{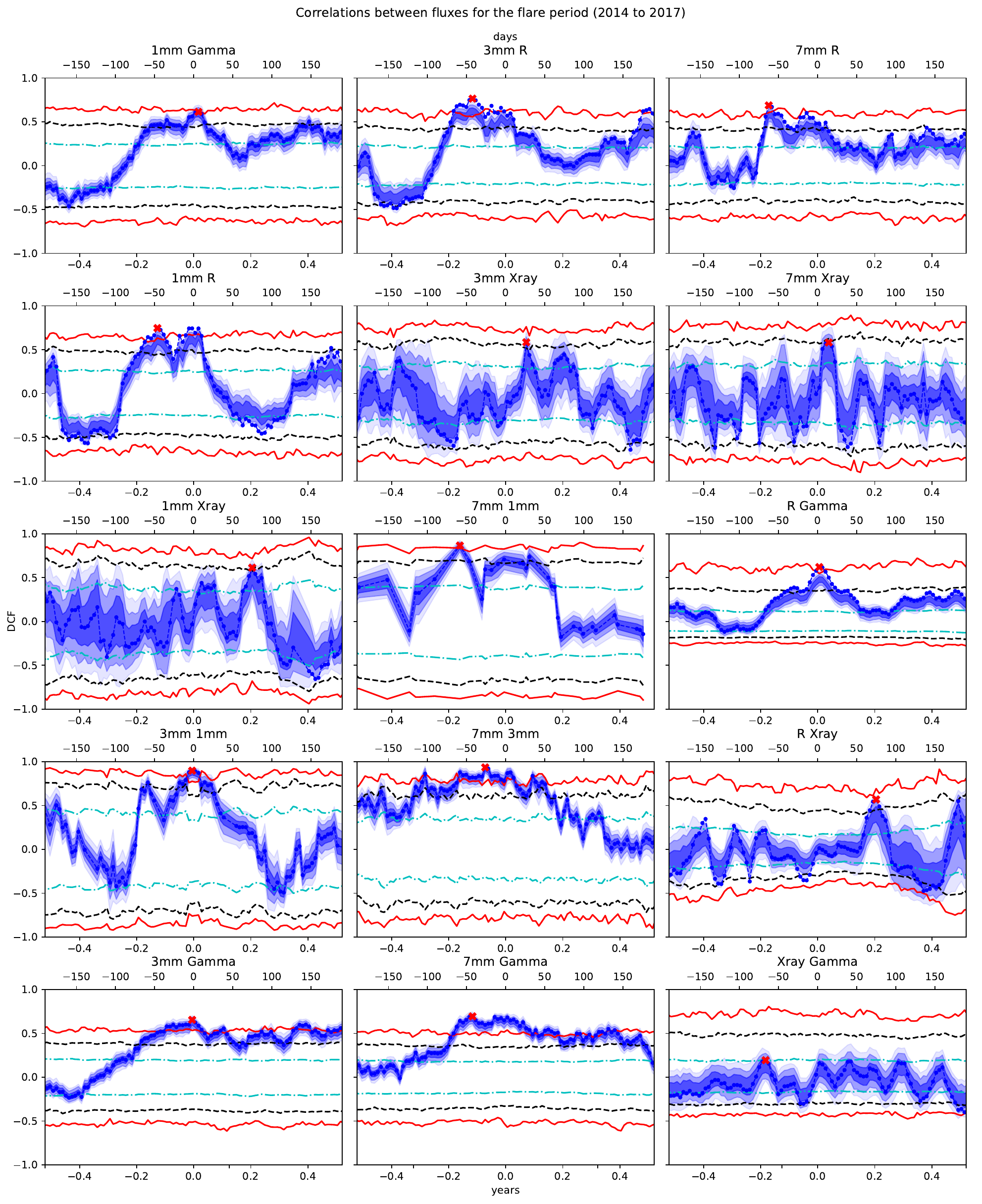}
            \caption{Correlations between fluxes across all wavelengths. Horizontal lines show significance levels for $1\sigma$, $2\sigma$, and $3\sigma$, computed using $N=2000$ synthetic light curves as described in Sec. \ref{sec:correlations}. The DCFs here are computed using only the flaring episode from 2014 to 2017.}
            \label{fig:2020_CLIP_2014_2017_corrs_flux}
        \end{figure*}

        The results from DCFs of the flaring episode (2014 to 2020) show again significant ($>3\sigma$) correlation between flux at all wavelengths, with most peaks positions close to zero (Fig. \ref{fig:2020_CLIP_2014_2017_corrs_flux}).
              
        The clear exception to this general correlation is again the X-ray band. The absence of $>3\sigma$ correlation close to zero for the X-ray emission with the other bands (Fig. \ref{fig:2020_CLIP_2014_2017_corrs_flux}) hints at other emission mechanisms, located at a different emission zone. This suggests that a different, separated processes could be responsible at least partially for the emission in X-rays, hypothesis also favored by the analysis of the spectral energy distribution of the source (section \ref{sec:seds}).

The interpretation of the peak positions in the DCFs is not straightforward. A debate on how accurately they represent the timing between different emission episodes is on-going. Specially as longer periods are taken into account, since more and different processes and regions can be involved in the correlation. For the DCF of the whole period,  the correlation only tells us about the probability that the processes causing the emissions are related. However, if we consider only the flaring episode, the relation between the peak position and timing of emissions will be more direct.
Even then, the presence of several correlation peaks makes the interpretation of the results difficult. These peaks are the consequence of the low, non-uniform sampling of available data, and the complex structure of the flares. This is a fundamental flaw of any correlation analysis, since this correlation noise might result in peaks that do not correspond to the real delay between the signals. Several ways of dealing with these have been proposed, such as using the centroid instead of the maximum, but they are not free from biases and flaws, such as those discussed in \textcite{Welsh:1999}. Here we follow \citeauthor{Welsh:1999}, and use simply the absolute maximum of the DCF, justifying the decision by the consistency of our results, as shown in the following.

        If the position of the peak is to represent the real delay between the signals at different wavelengths, these delays should be more or less compatible between themselves when computed using different sets of correlations, e.g. the delay between A and B plus the delay B and C should be close to the delay between A and C. In this sense it is possible to build a compatibility chart, showing the relations between the different positions. 
        
\begin{figure*}[hptb!]
            \centering
            \includegraphics[width=\textwidth]{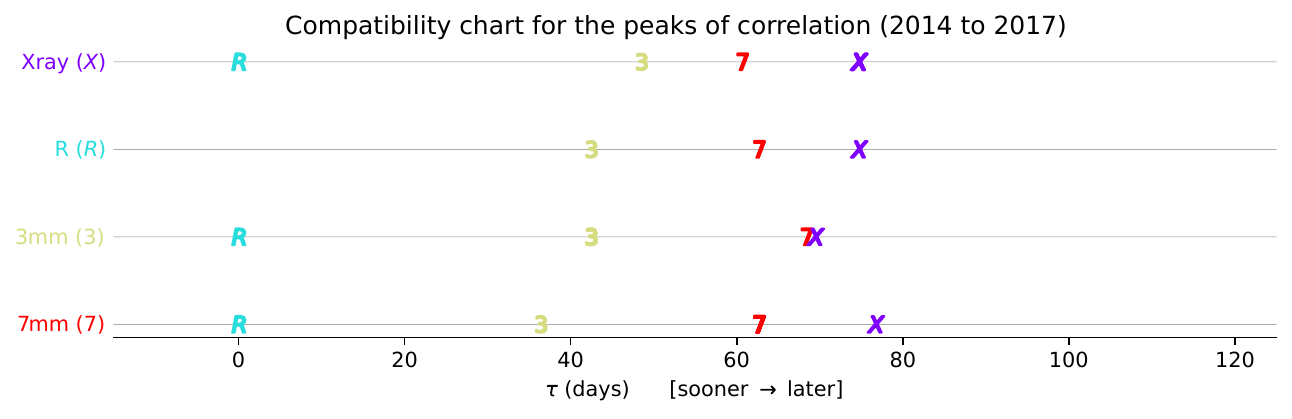}
            \caption{\lineskip=0pt Correlations compatibility chart for $\tau_p$. Each row corresponds to a band $i$. The delays or peak positions between the row band $i$ and any other band $j$, $\tau_p^{i,j}$, are plotted along the x-axis, shifted by the delay between the band $i$ and the reference band, $\tau_p^{i,R}$. The peaks selected are the one with highest values of the DCF, marked with red dots in Fig. \ref{fig:2020_CLIP_2014_2017_corrs_flux}. The DCFs here use the flaring episode from 2014 to 2017.}
            \label{fig:corr_I_compChart_tp}
        \end{figure*}

        This was done in Fig. \ref{fig:corr_I_compChart_tp} (Table \ref{tb:delay_table}) using the DCFs computed for the signals between 2014 and 2017 (the flare period, Fig. \ref{fig:2020_CLIP_2014_2017_corrs_flux}) and the band $R$ as a reference. Our choice of $R$ as the reference band is motivated by the fact that is the most densely sampled band during the periods of high variability.
In this graph, each row corresponds to a band $i$. The delays or peak positions between the row band $i$ and any other band $j$, $\tau_p^{i,j}$, are plotted along the x-axis, shifted by the delay between the band $i$ and the reference band, $\tau_p^{i,R}$, so that they fall aligned on the same positions.

        We see that indeed the positions fall more or less aligned in most cases, justifying our interpretation, and our choice of the maximum.
        The points for \si{1}{mm} and $\gamma$ are very dispersed, but the correlation between R and $\gamma$ presents a prominent peak, with high confidence and without spurious peaks close (Fig. \ref{fig:2020_CLIP_2014_2017_corrs_flux}), so we take $\tau_p^{\text{R},\gamma} \sim +\SI{2}{days}$ as the correct delay ${\tau}_{\text{R},\gamma}$.
        
\begin{figure}[htbp!]
            \centering
            \includegraphics[width=0.9\linewidth]{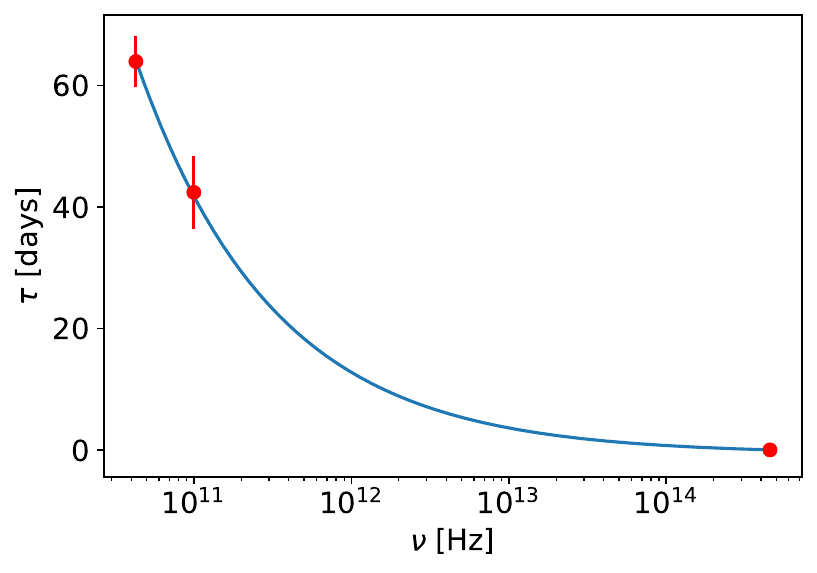}
            \caption{Fit to a function  $y = ax^{-1/2}+b$, which is the expected form if the time delays are due to different synchrotron cooling times. Fit results are: $a=\SI[scientific-notation = true, separate-uncertainty = true]{1.341(0.009)e+07}{}$, $b=\SI[scientific-notation = true, separate-uncertainty = true]{-6(1)e-01}{}$, $p(\chi^2)=\SI{0.909}{}$, $r^2=0.9998$. Parameter $b$ is close to zero and accounts for an arbitrary reference delay, R in our case.}
            \label{fig:delays_fit}
\end{figure}

        Discarding the \si{1}{mm} and Gamma rows, we estimate that the mean delays with respect to R for \si{7}{mm}, \si{3}{mm},  and X-ray, ${\tau}_{i,R} = \left\langle{\tau_p}^{i,R}\right\rangle$ are $\SI{64 \pm 4}{days}$, $\SI{42 \pm 6}{days}$  and  $\SI{73\pm 4}{days}$ respectively. The delays for \si{7}{mm}, \si{3}{mm}, and R are consistent with expectations if the mechanism of emission is synchrotron cooling, which should result in delays of the form $\tau_s \propto \nu_s^{-1/2}$, where $\nu_s$ is the synchrotron frequency, as can be seen in Fig. \ref{fig:delays_fit}.

        The delay obtained for X-ray with respect to R is much larger than for any other band. This strengthens the hypothesis that emission at X-ray energies might involve a different mechanism as already suggested by the lower level of correlation found, and as analysis of the spectral energy distributions in section \ref{sec:seds} reveal.

        \begin{table}[htbp!]
            \centering
\caption{Estimated delays (in days) obtained from peaks of the DCFs (Fig. \ref{fig:2020_CLIP_2014_2017_corrs_flux}) and represented in the compatibility chart (Fig. \ref{fig:corr_I_compChart_tp}), with their average and dispersion.
The delay between $R$ and $\gamma$ can be extracted directly from the DCF in Fig. \ref{fig:2020_CLIP_2014_2017_corrs_flux} and it is of \SI{2.0}{days}, as discussed in sec. \ref{sec:correlations}.}
\label{tb:delay_table}

\begin{tabular}{lrrrrrr}
    \toprule
    & 7mm & 3mm  & X-ray  \\
    \midrule
    7mm & - & 36.4  & 76.8   \\
    3mm & 68.7 & -   & 68.7  \\
R & 62.6 & 42.4   & 74.7  \\
    Xray & 60.6 & 48.5   & -  \\
\midrule
    mean  & 64.0 & 42.4  & 73.4  \\
    std & 4.2 & 6.1  & 4.2   \\
    \bottomrule
\end{tabular}
         \end{table}

    \subsection{Geometry of the emission regions}
    
        We can also compute the corresponding sizes implied by the variability timescales, since they are constrained due to causality and special relativity  according to the formulas
        \begin{equation}
            d = \dfrac{c \beta \Delta t }{(1-\beta\cos\theta)(1+z)}   
               = \dfrac{\beta_{app} c \Delta t}{(1+z) \sin\theta} 
\end{equation}    
        This same formula can also be used to compute the relative distances between emission regions implied by the time delays obtained in the correlation analysis, under the hypothesis that they result from the distances (although it needs not be the case as seen in the previous section if they arise from synchrotron cooling).
        
        The sizes of the emitting region can be constrained with
        \begin{equation}
            R_b = \dfrac{c  t_{var}  \delta_D}{1 + z}
            \label{eq:Rb_from_delta}
        \end{equation}
        For the moving component B2 corresponding to the 2008 outburst, using the timescale of variability  in Table \ref{tb:kinematics} and the $\delta_D \simeq 67$, one obtains sizes of around \SI{2}{pc}, consistent with the angular measure of VLBI images. Using the gamma-ray variability one obtains much lower sizes, of around \SI{0.2}{pc}, since variability at these energies is observed in timescales as short as \SI{8}{days} (\cite{Ackermann:2012}).
This smaller high-energy emitting region is in agreement with the expected result of synchrotron cooling in the proposed model which explains the longer duration of flaring activity in mm. The maximum viewing angle of this jet is cited to be $\lesssim \SI{2.4}{\degree}$ (\cite{Agudo:2011}) , which, through the relations between this angle and the true speed and Doppler factor
        \begin{align}
            \beta &= \sqrt{1-1/\Gamma^2}     \\ 
            \mu_s &= \dfrac{1}{\beta} \left(1 - \dfrac{1}{\Gamma\delta_D}\right) = \cos{\Theta}
        \end{align}
        and relation \eqref{eq:Rb_from_delta} limits to a minimum of \SI{1}{pc} the sizes of the mm-emitting regions.

        The relative distances of the core and knots to the base of the jet can be ascertained using a model for the geometry of the jet. Following \textcite{Wang:2020} and using a conical geometry,
        \begin{equation}
            r_{\text{core}} = \dfrac{r_{\perp}}{\varphi} = \dfrac{0.5 \theta_d d_L}{(1+z)^2 \varphi}			\,,
        \end{equation}
        where $\varphi$ is the half-opening angle of the jet and $\theta_d$ is the angular diameter.
        
        With our knot identification we can estimate the half-opening angle as $\varphi = (\Theta_{0,\text{max}}-\Theta_{0,\text{min}})/2 \simeq \SI{0.4}{\degree}$. However this way of estimating the half-opening angle is very sensitive to the weakest components. A second way to estimate this angle is (\cite{Weaver:2022}) $\varphi=\theta_{p}\sin{\Theta_{0}}$, where $\theta_p$ is the projected opening semi-angle of the jet and is taken to be twice the standard deviation of the jet position angle, or of the visible component in the case of a wobbling jet direction. With our parameters, this gives about \SI{0.78}{\degree}, closer to the more widely cited (\cite{Weaver:2022}, \cite{Wang:2020}) value of about  $\simeq \SI{1}{\degree}$ for B2, the brightest component and the responsible for the 2008 flare.

        For a core size at 43GHz of $\theta_d \simeq \SI{0.059}{mas}$ (similar to that obtained by \cite{Kutkin:2018}) this gives $r_\text{core,\SI{43}{GHz}} \simeq \SI{17}{pc}$. This would situate the distance from the base of the jet to the \SI{43}{GHz} core much closer than the  $r_\text{core,\SI{15}{GHz}} \simeq \SI{29}{pc}$ obtained by \textcite{Wang:2020} at 15GHz, consistent with opacity effects. The result is also compatible with the constraint $> \SI{12}{pc}$ from \cite{Agudo:2011}.

    \subsection{Spectral energy distribution} \label{sec:seds}
    
\begin{figure*}
            \centering
            \includegraphics[width=0.8\linewidth]{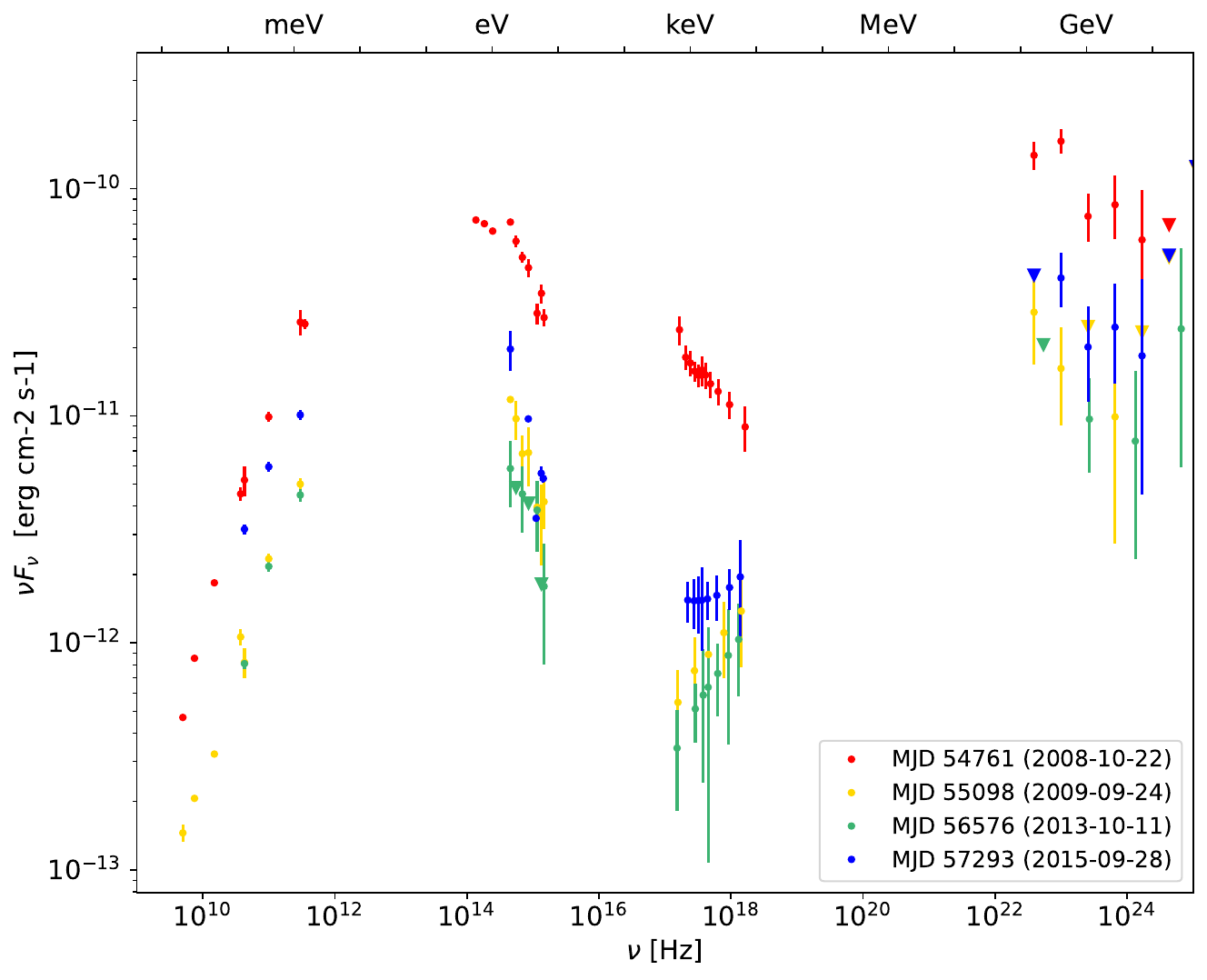}
            \caption{The four epochs for which the SEDs were analyzed, represented together for comparison. MJD 54761 and MJD 57293 correspond to flaring epochs of the 2008 and the 2015 outbursts respectively, while MJD 55098 and MJD 56576 correspond to quiescent epochs. The appearance of an X-ray bump is evident in the 2008 flaring epoch, and also visible, although dimmer, in the 2015 flaring epoch. Both quiescent epochs lack this X-ray feature.}
            \label{fig:seds_all_epochs}
        \end{figure*}

        We have produced complete SEDs for the two epochs of flaring and quiescent state related to the 2008 outburst where the MWL coverage was highest: MJD 54761 (2008-10-22), which corresponded to the peak of the flare, and MJD 55098 (2009-09-14). Analogously, we built the SED for two epochs related to the 2015 outburst: MJD  56576 (2013-10-11), which was taken as a quiescent epoch, and MJD 57293 (2015-09-28), as flaring epoch. The last epoch is the closest one to the peak of the flare with observations in enough bands to perform an accurate modeling. These four epochs were marked with vertical lines in the MWL flux plot (Fig. \ref{fig:2020_ALL_mwl_flux}) and their SEDs are represented together in Fig. \ref{fig:seds_all_epochs} for comparison.
        
\begin{figure}[htbp!]
            \centering
            \includegraphics[width=1\linewidth]{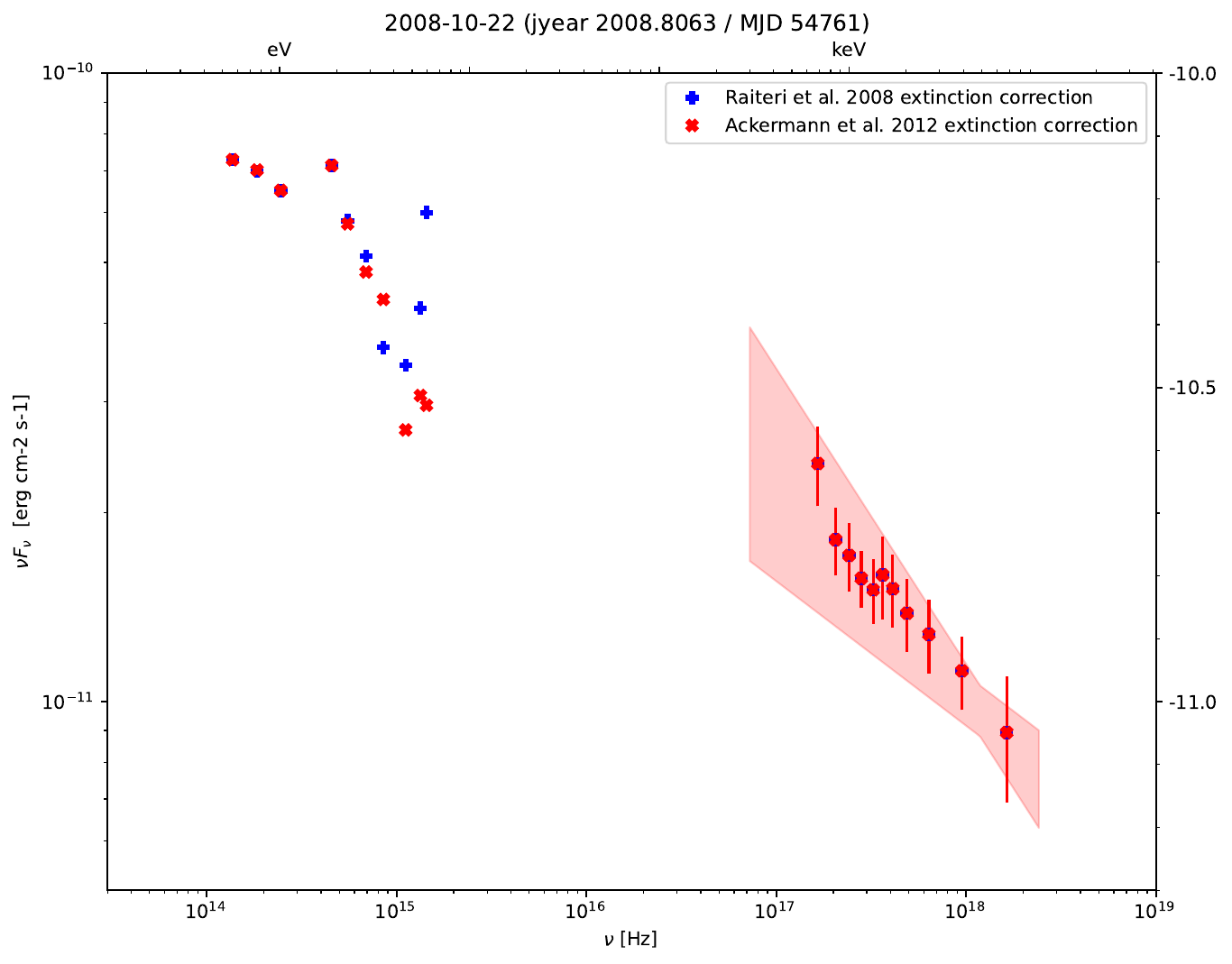}
            \caption{Resulting UV bump applying the extinction correction from \textcite{Raiteri:2008} and  \textcite{Ackermann:2012}. An increase is present in both cases but it is much dimmer with the values by  \textcite{Ackermann:2012}.}
            \label{fig:extcorr_comp}
        \end{figure}

        It can be seen that both the 2008 and 2015 flaring epochs, \hyperref[fig:sed_mjd54761]{MJD 54761} and \hyperref[fig:sed_mjd57923]{MJD 57923}, exhibit a softening of the spectrum between the hard UV and soft X-ray ranges (Fig. \ref{fig:seds_all_epochs}). The feature manifests itself as a increase of the flux from optical to UV wavelengths, with the slope in the SED in the UV becoming positive, and as a increased flux in the X-ray region, with the slope becoming negative. The UV increase is much higher when using the extinction values by \textcite{Junkkarinen:2004}, as seen in Fig. \ref{fig:extcorr_comp}, but this probably overestimates the correction in the hard UV. The feature is still present in both epochs when using the values for extinction given by \textcite{Ackermann:2012}, specially for MJD 57923 and considerably dimmer for MJD 54761, and we have used these values to built the final SEDs.  The unexplained feature seems to disappear when the source is quiescent, for both the 2008 and the 2015 flares.

        The origin of this feature is still under debate, although its presence has been reported before in the literature, also for previous flares of this source. \cite{Raiteri:2008} reported the presence of the UV feature in the peak of the 2006-2007 flare, and also in some other earlier epochs where the source was fainter. The change of slope in X-rays was also present in the SED reported in \textcite{Ackermann:2012} for the MJD 54761-3 epoch, even though the UV increase was not evident in their SED plots. In contrast, our analysis shows that in epoch \hyperref[fig:sed_mjd54761]{MJD 54761} that the bump is visible both in the UV and the X-ray.
\textcite{Raiteri:2008} emphasizes that the fact that the bump is visible during flaring states is unusual for quasars. In most cases, similar features are only visible in the faintest states and are attributed to thermal emission from the disk. In contrast, 0235+164 exhibits this feature even during the brightest epochs, hence ruling out such an explanation. The thermal origin of the feature is further discarded by the high temperatures that would be necessary to result in a bump at these energies, and by the fact that the thermal emission from the disk should be approximately stable, while the difference in flux when the feature becomes visible is of more than one order of magnitude. \textcite{Raiteri:2008} also reported the presence of the feature in the UV for a quiescent epoch related to the 2007 flares, and some intermediate states. This, together with the lower values for the correlation of the X-rays found in our DCF analysis (section \ref{sec:correlations}), hints at a different process involved at least partially in the emission at these energies.

\begin{figure}[htbp!]
            \centering
            \includegraphics[width=1.0\linewidth]{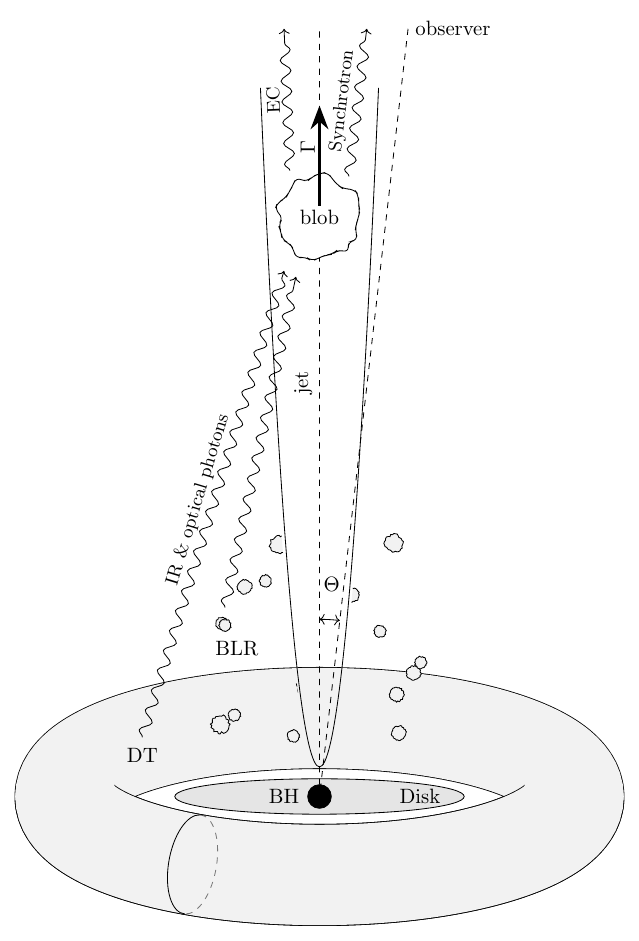}
            \caption{Schematic representation of the physical setup. In the proposed scenario, an electron distribution (blob) in the jet, characterized by its geometrical properties, its energy distribution and its magnetic field (Tables \ref{tb:sed_models_ec_plc} and \ref{tb:sed_models_ssc_plc}), emits Synchrotron radiation. Infrared and optical photons from the Disk, the Dusty Torus and the Broad Line Region reach the blob and are up-scattered to high energies by Inverse Compton. The observer, narrowly aligned with the jet, sees the emission boosted by relativistic effects. In the case of Bulk Compton emission, an additional, different distribution would exist, closer to inner region of the blazar (Table \ref{tb:sed_models_bc}).}
            \label{fig:sketch_blazar_model}
        \end{figure}

        In the remaining of this section we will briefly review previous existing models and perform a comparison between them and ours, the result of which is summarized in Table \ref{tb:model_review_summary}. A schematic representation of the physical setup can be found in Fig. \ref{fig:sketch_blazar_model}.
        
\begin{table*}
            \centering
\caption{Summary of the comparison between different models for the flaring epoch MJD 54761.}

\begin{tabular}{lrccc} \toprule
Reference & 
        & \textcite{Agudo:2011} 
        \tablefootmark{b}
        &  \textcite{Ackermann:2012} 
                \tablefootmark{c}                   
        &  \textcite{Baring:2017} 
                \tablefootmark{d}
        
\\ 
\midrule
\multirow{2}{*}{Model components\tablefootmark{a}} & 
        &  \multirow{2}{3cm}{\centering Synch + SSC (dom.)} 
        &  \multirow{2}{3cm}{\centering Synch + EC (DT) (dom.) + SSC + BC} 
        &  \multirow{2}{3cm}{\centering Synch + EC (BC) (dom.) + SSC } 
\\
\\
\midrule
Bulk Lorentz factor & $\Gamma$ 
        & 39.4
        &  20 ($\Gamma_b = 10$) 
        &  35 
\\
\midrule
Viewing angle  & $\Theta$ 
        & $\leqq \ang{2.4}$                       
        &  \ang{2.3} 
        &  \ang{1.7} 
\\
\midrule
Doppler factor  & $\delta$ 
        & 24                    
        &  20 ($\delta_b = 16$) 
        & -
\\
\midrule
Opening angle  & $\theta$ & \ang{2.3}            
        & \ang{2.9} ($\theta_{b}=\ang{2.3}$) 
        & -
\\
\midrule
Location & $r$  
        & \SI{12}{pc}             
        & \SI{1.7}{pc}  
        & -
\\
\midrule
Size of the emission region & $R$ 
        & -   
        & -
        & \SI{1e16}{cm}
\\
\midrule
Magnetic field intensity & $B$ 
        & -              
        & \SI{0.22}{G} 
        & \SI{2.5}{G} 
\\
\midrule
\multirow{1}{*}{Electrons} 
        & $\gamma_{\text{min}}$ 
        &  \multirow{2}{*}{-}                    
        & \SI{100}{} 
        & - 
\\
        & $\gamma_{\text{max}}$ 
        &                    
        & \SI{5.8e3}{} 
        & \SI{1.61e3}{} 
\\
        & \makecell{$n_{e^-}(E)$ \\ ~} 
        & -
        & \makecell{doubly broken pwl. \\ $p_1 = 1.5$, $p_2 = 2.03$, $p_3 = 3.9$} 
        & See foot note\tablefootmark{e}
\\
\midrule
Protons 
        & $n_{e^-}/n_{p^+}$
        & -
        & 9
        & -
\\
\midrule
Disk luminosity & $L_{\text{disk}}$ 
        & - 
        & \SI{4e45}{erg s^{-1}} 
        & \SI{3.4e44}{erg s^{-1}} 
\\
\midrule
Disk temperature & $T_{\text{disk}}$  
        & -      
        & \SI{3.5e3}{K}\footnote{From the given radiation temperature of \SI{0.3}{eV}.}    
        & \SI{1e3}{K}  
\\
\midrule
Disk radius & $R_{\text{disk}}$ 
& -
& -
& \SI{6e17}{cm}
\\
\bottomrule
\end{tabular}

\tablefoot{
\\
\tablefoottext{a}{The components considered are Synchrotron (Synch), Synchrotron Self-Compton (SSC), External Compton (EC) from the Dusty Torus (DT), and Bulk Compton (BC).}
\\
\tablefoottext{b}{The values for \textcite{Agudo:2011} do not come from a SED model, but from the DCF analysis  and kinematic parameters from VLBI images, assuming a SSC  scenario. The value for $\Gamma$ is not cited in the paper, it is the one obtained by \cite{Weaver:2022} for the same component in VLBI.} 
\\
\tablefoottext{c}{Parameters for the blazar zone (their model includes a second population of relatvistic cold electrons to account for the secondary soft X-ray bump, whose parameters are indicated between parenthesis).} 
\\
\tablefoottext{d}{The secondary soft x-ray bump is modeled by bulk Comptonization of a background seed field from a dusty torus); the H.E. bump by External Compton of the electron population.} 
\\
\tablefoottext{e}{The energy distribution is simulated from Diffusive Shock Acceleration (DSA), resulting parameters are not explicitly indicated.}
}             \label{tb:model_review_summary}
        \end{table*}

        \textcite{Agudo:2011} postulated that the mechanism of emission was predominantly SSC from the joint analysis of VLBI images, long-term multi-wavelength light curves from \SI{}{mm} to $\gamma$-ray energies including polarization, and time delays. 
They interpreted the outburst as "a consequence of the propagation of a disturbance, elongated along the line of sight by light-travel time delays, that passes through a standing recollimation shock in the core and propagates down the jet to create the superluminal knot". 
They also demonstrated the general correlation between the MWL flux at different bands and the appearance of the \SI{43}{GHz} VLBA superluminal features, and obtained the associated time delays. They argued that the variability in $\gamma$ rays could not be explained within the EC scenario. Instead, they favored a model where the stronger variability in $\gamma$ rays is explained by the delayed variability in a multi-zone turbulent cell model (\cite{Marscher:2010}). This was supported by the general multi-wavelength correlation, the variability of the polarization and the parameters derived from the superluminal components in VLBI images (see Table \ref{tb:model_review_summary}).

        \textcite{Ackermann:2012} produced a model of the SEDs for epochs 54761-3 (2008-10-22 - 24) and 54803-5 (2008-12-03 - 05). The high state epoch 54761 presented a secondary  soft X-ray bump which was modeled as a bulk-Compton feature, although no hint of a bump was present in the hard UV region in the SED.
For both of the epochs, ERCIR (Compton emission from Infrared radiation from the dusty torus) was the dominating component at higher frequencies. The bulk-Compton feature was not present in the quiescent state.
\citeauthor{Ackermann:2012} argues that EC must dominate SSC for any reasonable covering factor of the broad-line region. The model presents an emission zone located outside the BLR close to the BH ($\SI{1.7}{pc}$) with a Lorentz factor $\Gamma=20$, opening angle $\SI{2.9}{\degree}$, magnetic field $B^\prime=\SI{0.22}{G}$ and viewing angle $\SI{2.3}{\degree}$. The electron energy distribution was modeled by a doubly-broken power-law. The bulk-Compton feature is modeled by a population of cold electrons.

        \textcite{Baring:2017} models the same epoch MJD 54761. They do so with a Lorentz factor $\Gamma = 35$. They  model the energy distribution of electrons by simulating their acceleration process through Diffusive Shock Acceleration (DSA). The bulk Compton feature is also not present in the quiescent state. First and second order SSC also contributes to the second bump but is dominated at all energies by the external Compton. However, the authors notice that the Lorentz factor required by this source is significantly higher than the usual ($\Gamma \sim 10-20$) for EC-dominated sources. The Swift-XRT excess is modeled as IC of a seed radiation field of $T \sim \SI{1000}{K}$, postulated to be a dusty torus.
\textcite{Dreyer:2021} also presented a modeling of the SED for the same epoch, based on \textcite{Baring:2017}, where the X-ray bump is explained by bulk-Compton emission. The second bump is also explained by external Compton from the dusty torus.
If bulk Comptonization is responsible for the X-ray bump, a prediction is made that it should result in partial polarization in the X-ray bands.

\begin{figure}[hptb!]
            \centering
            \includegraphics[width=1\linewidth]{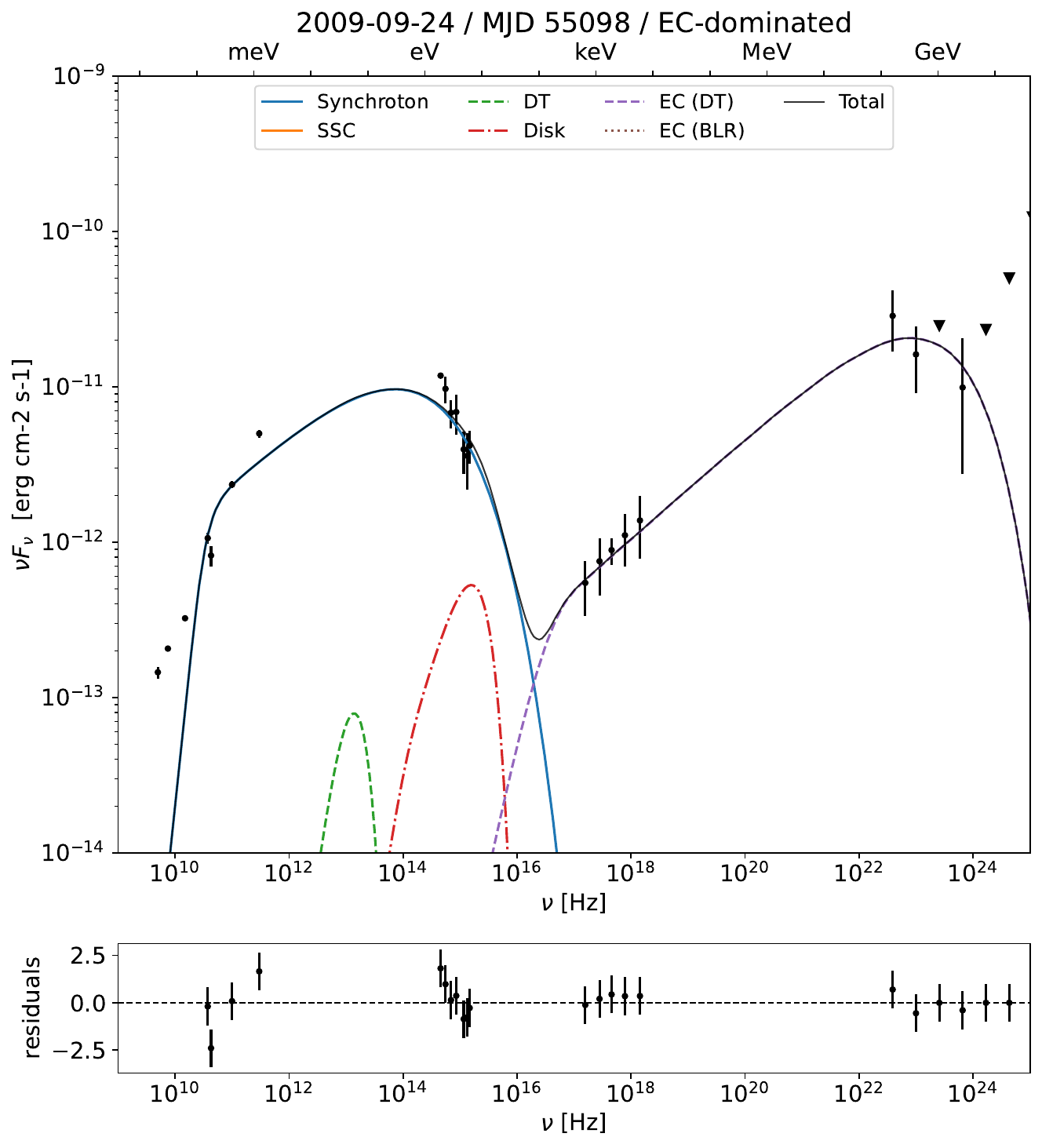}
            \caption{SED model for epoch MJD 55098 in the external Compton scenario (Table \ref{tb:sed_models_ec_plc}).}
            \label{fig:sed_mjd55098_ec_plc}
        \end{figure}

\begin{figure}[hptb!]
            \centering
            \includegraphics[width=1\linewidth]{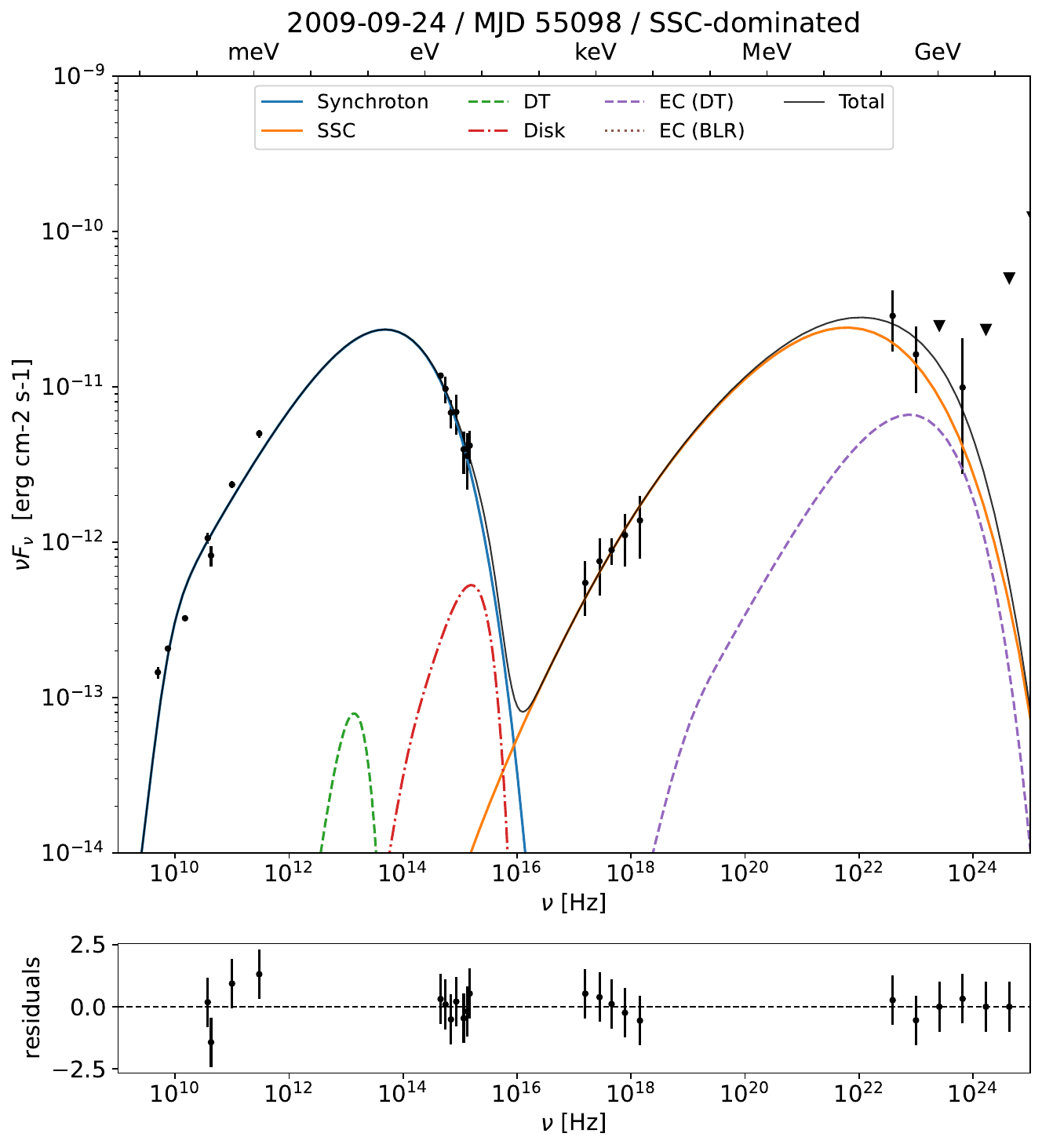}
            \caption{SED model for epoch MJD 55098 in the SSC-dominated scenario (Table \ref{tb:sed_models_ssc_plc}).}
            \label{fig:sed_mjd55098_ssc_plc}
        \end{figure}

\begin{figure}[hptb!]
            \centering
            \includegraphics[width=1\linewidth]{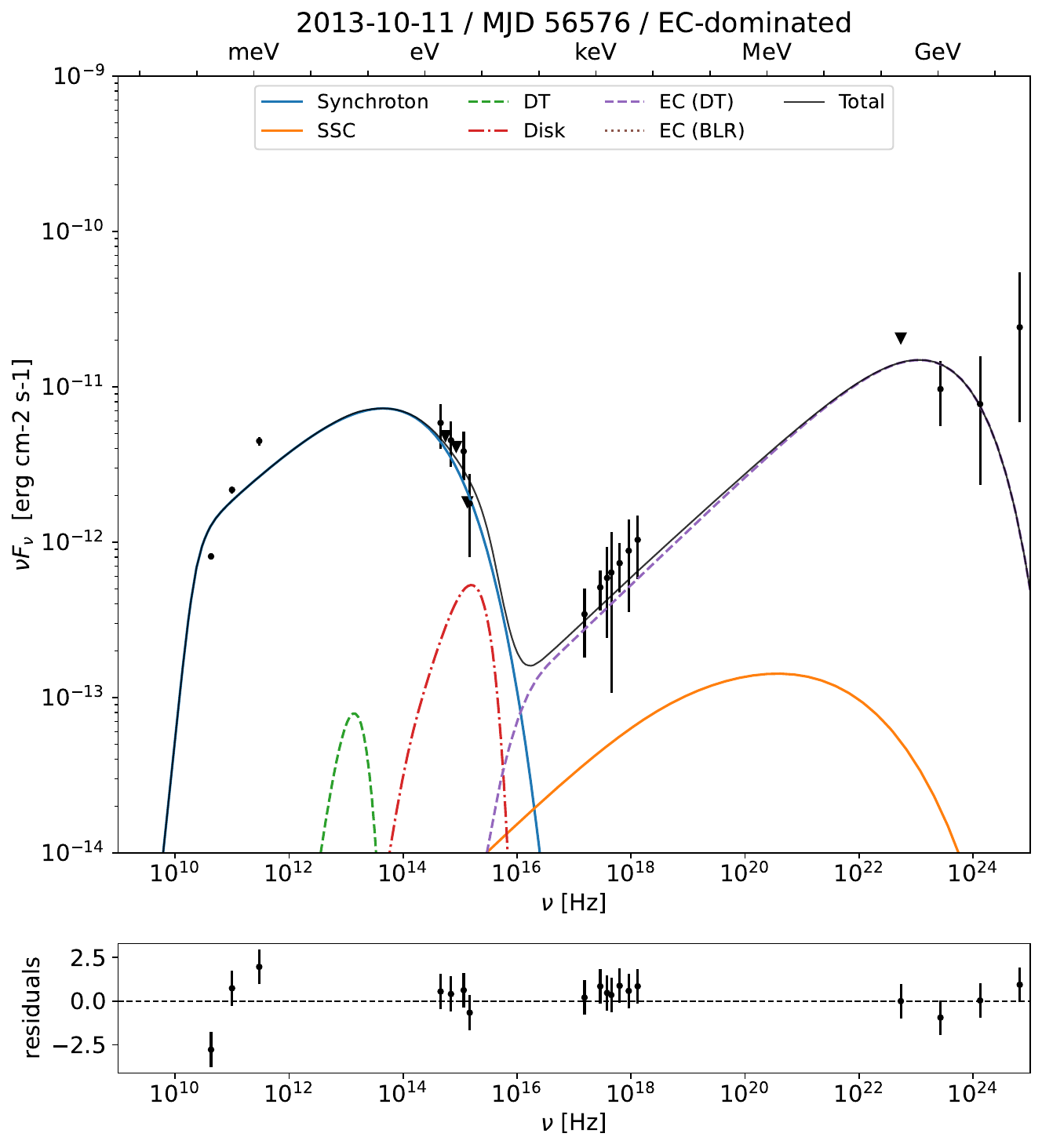}
            \caption{SED model for epoch MJD 56576 in the EC-dominated scenario (Table \ref{tb:sed_models_ec_plc}).}
            \label{fig:sed_mjd56576_ec_plc}
        \end{figure}

\begin{figure}[hptb!]
            \centering
            \includegraphics[width=1\linewidth]{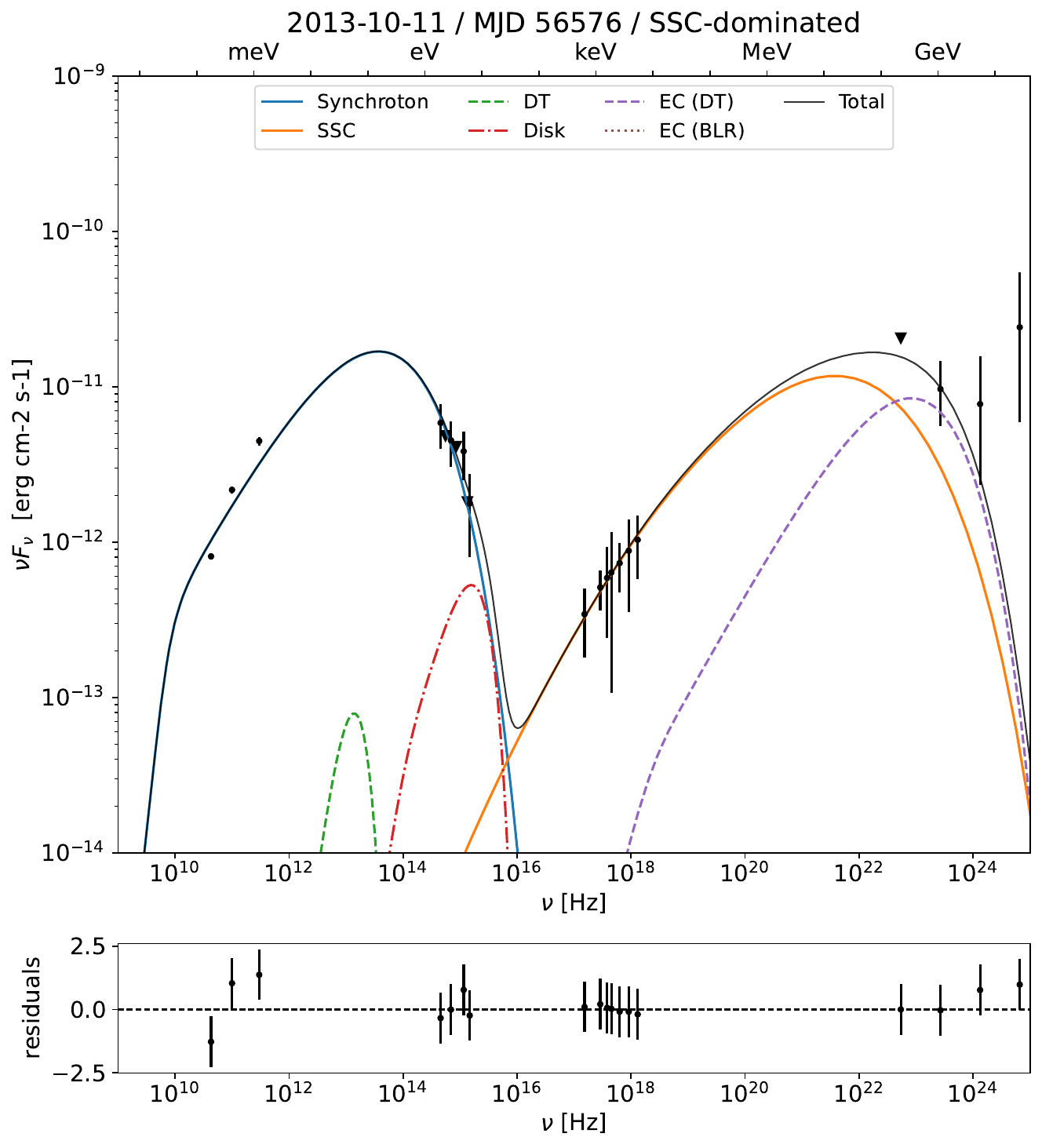}
            \caption{SED model for epoch MJD 56576 in the SSC-dominated scenario (Table \ref{tb:sed_models_ssc_plc}).}
            \label{fig:sed_mjd56576_ssc_plc}
        \end{figure}

        In this work, we have modeled the emission of AO 0235+164 using the \texttt{JetSeT framework}\footnote{\url{https://jetset.readthedocs.io/en/1.1.2/}} framework (\cite{Tramacere:2020,Tramacere:2011,Tramacere:2009}), using a SSC + EC scenario. The accretion disk spectrum is modeled as a multi-temperature black body as described in \cite{King:2002}, with a luminosity fixed to $L_{\rm Disk}=5\times 10^{45}$, erg/s, an accretion efficiency ($\eta$) fixed to the standard value of 0.08, and a BH mass fixed to $5\times 10^8 M_\odot$, with an external radius of the order of a few hundreds of Schwarzschild radii.
        The BLR is modeled as a thin spherical shell with an internal radius determined by the phenomenological relation provided by \cite{Kaspi:2007},  $R_{BLR,in}=3\times10^{17}L_{\rm Disk,46}^{1/2}\,$cm. The external radius of the BLR is assumed to be $0.1 R_{\rm BLR,in}$, with a coverage factor $\tau_{BLR}=0.1$. The dusty torus (DT)  is assumed to be described by spherical uniform radiative field, with a radius $R_{DT}=2\times10^{19}L_{\rm Disk,46}^{1/2}\,$cm,  \citep{Cleary:2007}, and a reprocessing factor $\tau_{DT}=0.1$. The emitting region is modeled  as a single spherical zone with a radius $R$, located at a distance $R_{H}$ from the central black hole. The jet has  a conical geometry, with an half opening angle  $\phi\approx$ 3 deg, with the emitting region size determined by  $R=R_{H} \tan{\phi} $. The emitting region moves along the jet axis with a bulk Lorentz factor $\Gamma$, oriented at a viewing angle $\theta$, and a consequent beaming factor $\delta= 1/(\Gamma \sqrt{1-\beta_{\Gamma} \cos(\theta))}$. For  relativistic emitting electron distribution (EEE) we tested a  broken power law (BKN) distribution
         \begin{equation}
             n(\gamma)= N \left\lbrace
             \begin{array}{ll}
              \gamma^{-p} & \gamma_{min}\leq \gamma \leq \gamma_{b} \\
             \gamma^{-p_1}\gamma_{b}^{p-p_1} & \gamma_{b}<\gamma<\gamma_{max},
             \end{array}
             \right.
             \label{eq:bkn}
         \end{equation}
        with an index of $p$ and $p_1$ below and above the break energy $\gamma_b$, respectively, and a powerlaw distribution with a cut-off (PLC) 
        distribution
         \begin{equation}
             n(\gamma)= N  \gamma^{-p} \exp{\frac{\gamma}{\gamma_{\rm cutoff}}},   \gamma_{min}\leq \gamma \leq \gamma_{max}            
             \label{eq:plc}
         \end{equation}
         The initial values of $L_{\rm Disk}$ and $T_{\rm Disk}$, are determined by \texttt{JetSeT} during the pre-fit stage, and $L_{\rm Disk}$ is frozen to the value of   $L_{\rm Disk}=5\times10^{45}$ erg s$^{-1}$.  The  model minimization is performed using the \texttt{JetSeT} \texttt{ModelMinimizer} module plugged to \texttt{iminuit} python interface \citep{iminuit}. The errors are estimated from the matrix of second derivatives, using the \texttt{HESSE} method. 
We fit data above 30 GHz, excluding data below the synchrotron self-absorption frequency.  To avoid that the small errors in the UV-to-radio frequencies biasing the fit toward the lower frequencies, we add a 20\% systematic error to data below $10^{16}$ Hz.
         We find that the PLC model provides a slightly better fit to the data, hence in the following we present only the results for this model. All the states presented in this analysis can be modeled by a single-zone EC-dominated (see Figures : \ref{fig:sed_mjd55098_ec_plc}, \ref{fig:sed_mjd56576_ec_plc}, \ref{fig:sed_mjd57293_ec_plc_no_bulk} and  Table \ref{tb:sed_models_ec_plc}) or SSC-dominated scenario (see Figures: \ref{fig:sed_mjd55098_ssc_plc},\ref{fig:sed_mjd56576_ssc_plc} and   Table \ref{tb:sed_models_ssc_plc}), with the SSC-dominated scenario resulting in systematically lower values of $B$, needed to accommodate for the proper $U_e/U_B$ ratio able to match the peak flux and frequency of the IC emission. On the contrary, for the flaring state on MJD 54761, the presence of a strong and soft bump in the X-ray makes  both the SSC and EC  unable to model the data. As suggested by \cite{Celotti_Ghisellini_Fabian_2007, Ackermann:2012}, this spectral feature can be explained by the Comptonization of the external radiative fields by a population of cold electrons. We have introduced such bulk Compton (BC) component, modeled as 
         a spherical region with a radius $R_{BC}$ moving with corresponding bulk factor $\Gamma=10$, at a distance of $r$ from the BH, and with a total number of particle $N_{BC}$.

\begin{figure}[hptb!]
            \centering
            \includegraphics[width=1\linewidth]{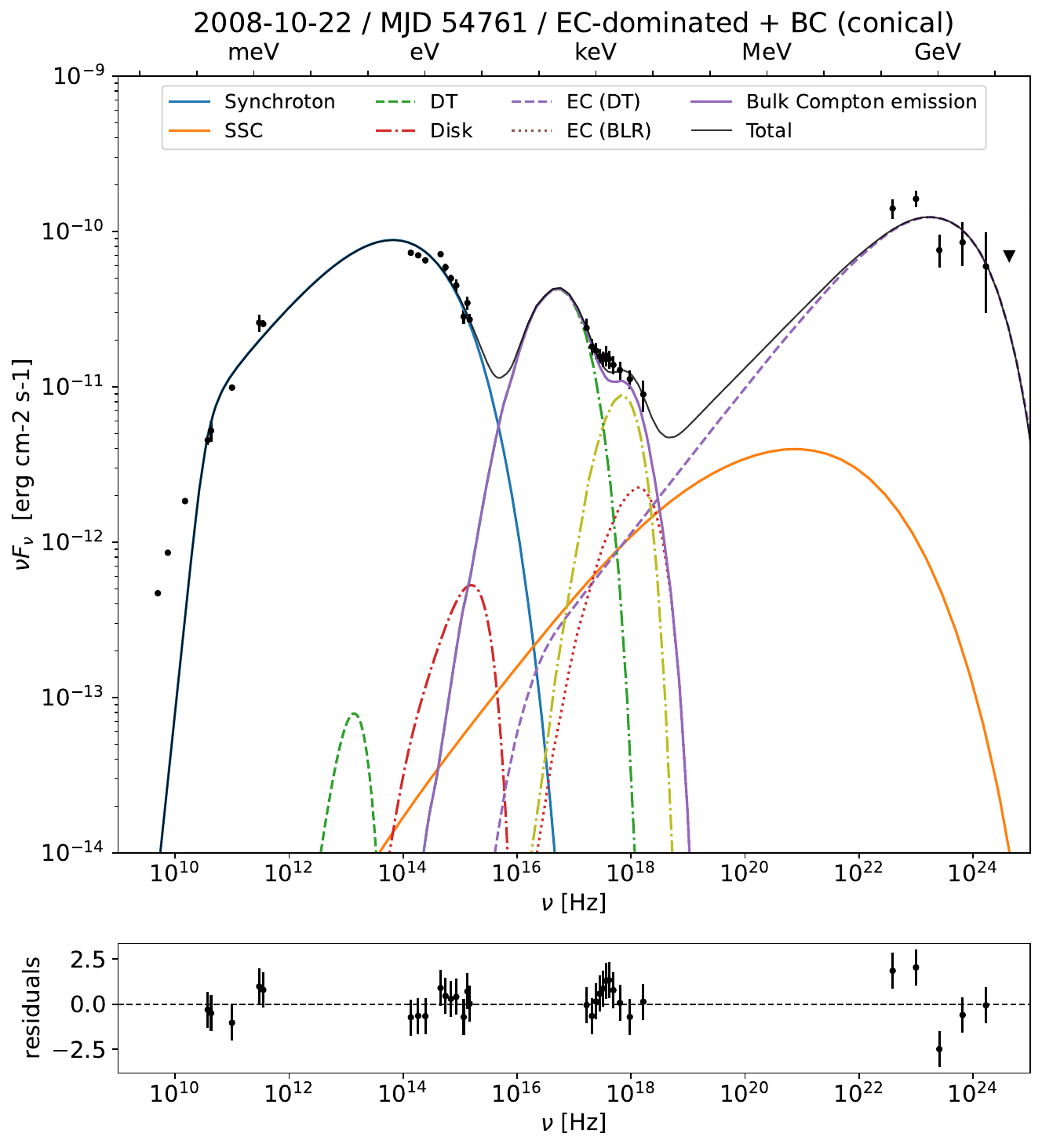}
            \caption{SED model for epoch MJD 54761. The EC-dominated model includes a bulk Compton component with a conical shape.}
            \label{fig:sed_mjd54761_ec_w_bulk_conical}
        \end{figure}

\begin{figure}[hptb!]
            \centering
            \includegraphics[width=1\linewidth]{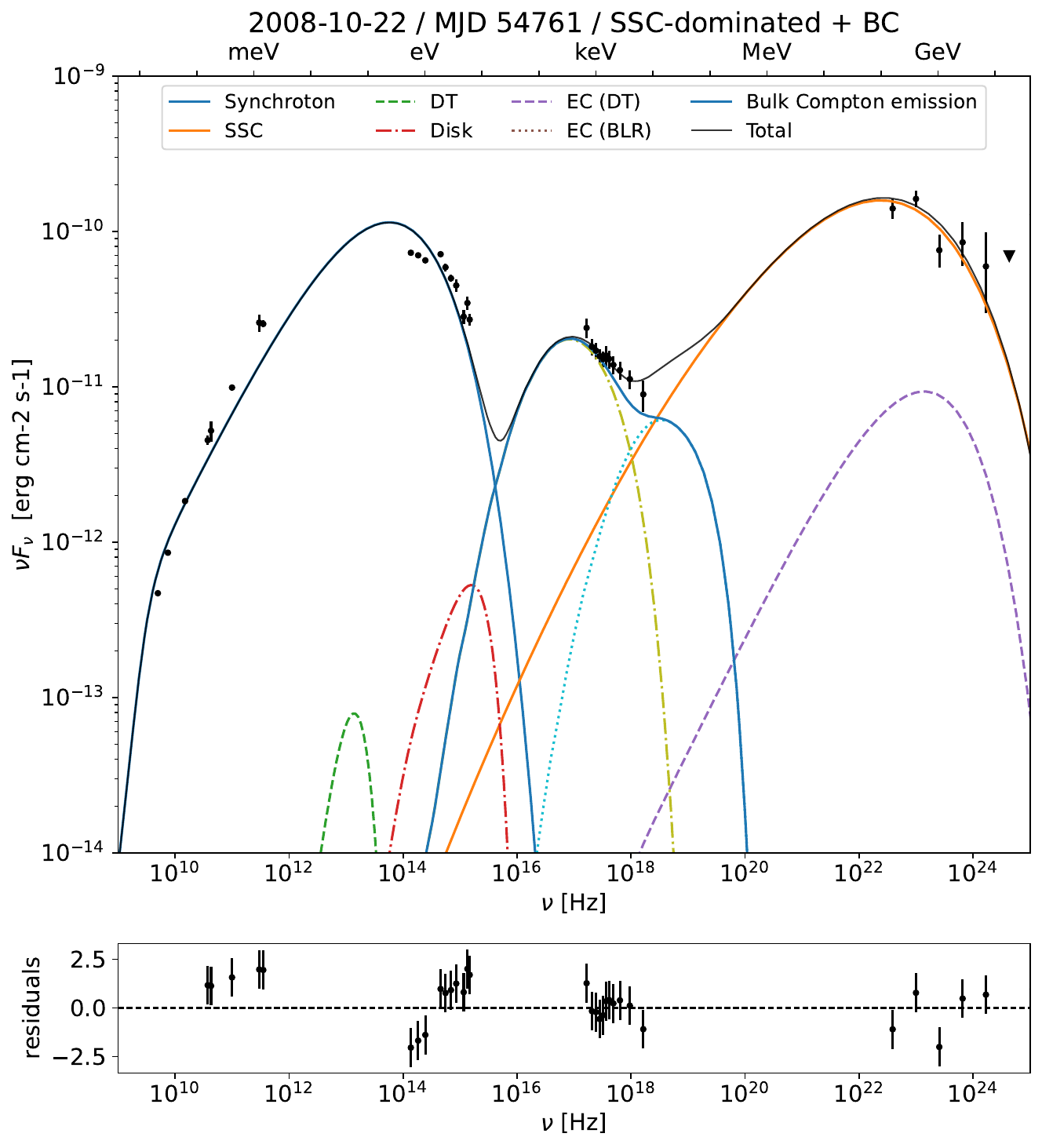}
            \caption{SED model for epoch MJD 54761 in the SSC-dominated scenario. The X-ray bump is modeled as bulk Compton emission from the Disk in a similar way to the EC-dominated model (Table \ref{tb:sed_models_bc}).}
            \label{fig:sed_mjd54761_ssc_w_bulk}
        \end{figure}

         We noticed that  for a purely cold population, i.e. for electron with  $\gamma_{min}=\gamma_{max}=\gamma=1$, the resulting shape of the BC radiation was always to steep to reproduce the observed data (see e.g. \cite{Celotti_Ghisellini_Fabian_2007}), on the contrary, we found that  a reasonable fit to the data was provided by increasing the fit range of  $\gamma_{max}$ to \num{5}, and setting $r=\SI{1.5e16}{cm}$. With this model configuration, the fit converged with a resulting value of $\gamma_{max} \approx \num{4}$ and  a resulting total number of cold electrons $N_{BC} \approx \num{1.8e54}$ (see Figure \ref{fig:sed_mjd54761_ec_w_bulk_conical} and left column in Table \ref{tb:sed_models_bc}). These values are compatible with those reported in \cite{Ackermann:2012} ($N_{BC}=\num{2.4e54}$ and  $r=\SI{5e15}{cm}$), anyhow we stress that in  \cite{Ackermann:2012} the BC spectral shapes is assumed to be a PL, whilst, in the present analysis it is obtained by the actual  Comptonization of the  cold electrons. 
          We also applied the BC model to an  SSC-dominated scenario, (see Fig. \ref{fig:sed_mjd54761_ssc_w_bulk} and Tables \ref{tb:sed_models_bc} and \ref{tb:sed_models_ssc_plc}), we notice that even though the overall agreement of the model with the data is still reasonable, the model shows a tension, in the optical-IR and X-ray data. At the high-energy branch of the X-ray data,  the excess of flux in the model is due to the broader spectrum of IC emission compared to the EC case, originating to the broader spectrum of the synchrotron seed photons compared to the narrower seed photon spectrum of the external fields. 
         
         Another possible choice, able to produce a PL shape of the BC, can be obtained assuming a purely cold electron population  with a truncated conical geometry, with the higher energy part of the BC being produced by the low-number electrons closer to  BH, and the higher energy  being produced by the larger number of electrons in the upper part of the truncated cone. To mimic such a geometry we implemented a BC model with two spherical regions.The radius of the two regions is obtained in order that the two spheres match the volume of the upper and lower part of the truncated cone. We find that a reasonable modeling of the BC emission is obtained assuming a truncated cone, with an opening angle of $45 ^\circ$ and an height of $\approx\SI{9E15}{cm}$, with the smaller  spherical region corresponding to the segment of the truncated cone  with an height of $\approx\SI{5E15}{cm}$, and the larger spherical region corresponding to the segment with an height of $\approx \SI{8E15}{cm}$ The total number of cold electrons is of $N_{BC} \approx \SI{1.4E54}{}$ (see Figure \ref{fig:sed_mjd54761_ec_w_bulk_conical} and left column in Table \ref{tb:sed_models_bc}). 
         Since the the introduction of this extra component introduces new parameters, first, we used the \texttt{ModelMinimizer} to fit the model to the  data without the BC component and excluding the X-ray data (the statistics are reported in  Table \ref{tb:sed_models_ec_plc}, and \ref{tb:sed_models_ssc_plc}), and in a second step, we added the X-ray data and we proceeded to a qualitative fitting of the BC conical component (the values of BC component are reported in Table \ref{tb:sed_models_bc}).
        
The flaring epoch MJD 57293 could also be modeled using a single zone model, although the observed softening of the X-ray spectrum could be explained by bulk Compton emission in two-zone model, in a similar manner to MJD 54761, as the DCF analysis for the 2014-2017 points towards to.

        \begin{table*}[htbp!]
            \centering
\caption{Parameters for the models of the Bulk Compton (BC) emission both with simple and conical geometries, shown in Figs. \ref{fig:sed_mjd54761_ssc_w_bulk}, \ref{fig:sed_mjd54761_ec_w_bulk}, \ref{fig:sed_mjd54761_ec_w_bulk_conical} for epoch MJD 54761.}

\begin{tabular}{lrlrrr} \toprule
Epoch & &
        & MJD 54761
        & MJD 54761
        & MJD 54761
\\ 
\midrule
Model \tablefootmark{a} & &
    & \makecell[r]{SSC-dominated +\\ BC (simple)} 
    & \makecell[r]{EC-dominated +\\ BC (simple)} 
    & \makecell[r]{EC-dominated +\\  BC (conical)}

\\
    \midrule
    \multicolumn{2}{l}{Geometrical parameters} & 
    \\[0.2em]
        ~ Bulk Lorentz Factor & $\Gamma$ &
            & \num{10}
            & \num{10}
            & \num{10}
        \\
        ~ \makecell[lt]{ Location \\ ~ (extended)} & $r$ & \si{pc}
            & \makecell[tr]{ \num{4.86e-3}   \\ -}
            & \makecell[tr]{ \num{4.86e-3}  (\num{<1e-1}) \\ - }
            & \makecell[tr]{ \num{1.30e-4}  \\ \num{2.76e-3} }
        \\
        ~ \makecell[lt]{ Size \\ ~ (extended)} & $R$  & \si{pc}
            & \makecell[tr]{ \num{3.28e-3}  (\num{3.83e-5})	\\ -}
            & \makecell[tr]{ \num{3.24e-3}   \\ -}
            & \makecell[tr]{ \num{1.56e-4}   \\ \num{3.32e-3} }
        \\
        ~ \makecell[lt]{Light crossing time \\ ~ (extended)}  &  $t^{\mathrm{obs}}_{\mathrm{var}}(R,\Gamma,\theta)$  & \si{day}
        & \makecell[tr]{ \num{0.4} \\ - }
        & \makecell[tr]{ \num{0.4} \\ - }
        & \makecell[tr]{ \num{0.02}  \\ \num{0.4}}
   \\
   \midrule
   ~ \makecell[lt]{ Magnetic field \\ ~ (extended)} & $B$  & \si{G}
       & \SI{9.14e-02}{G}  (\num{4e-02})
       & \SI{9.19e-02}{G}  (\num{7e-03})
       & \makecell[tr]{	\num{0.1} \\ \num{0.1}}
    \\
    \midrule
    \multicolumn{2}{l}{Particle distribution} & 
    \\[0.2em]
        ~ Minimum Lorentz factor & $\gamma_{\text{min}}$ &
            & \num{1.0}
            & \num{1.00}
            & \num{1.0}
        \\
        ~ Maximum Lorentz factor & $\gamma_{\text{max}}$ &
            & \num{4.0} (\num{1e-01})
            & \num{4.33} (\num{<1e-2})
            & \num{1.2}
        \\
        ~ Type & $n_{e^-}(E)$ &
            & PL
            & PL
            & PL        
        \\
        ~ Density & $N$ & \si{cm^{-3}}
            & \num{4.46e+05} (\num{2e+04})
            & \num{4.33e+05} (\num{<1e-3})
            & \num{3.11e+05}
        \\
        ~ Spectral slope & $p$ &
            & \num{2.85} (\num{7e-02})
            & \num{3.00} (\num{<1e-2})
            & \num{1.0}
\\
\bottomrule
\end{tabular}

\tablefoot{
    Uncertainties for the best-fit values were automatically obtained using the \texttt{HESSE} method of second derivatives and are indicated between parenthesis, parameters without them were frozen during the fit.  For uncertainties smaller than the third significant digit, an upper limit is given. The rest of the parameters for the models can be found in Tables \ref{tb:sed_models_ec_plc} and \ref{tb:sed_models_ssc_plc}, together with their uncertainties and fit statistic.
    \\
    \tablefoottext{a}{Only the parameters of the Bulk Compton emission are shown here. See Tables \ref{tb:sed_models_ec_plc} and \ref{tb:sed_models_ssc_plc} for the rest of the parameters.}
}             \label{tb:sed_models_bc}
        \end{table*}
    
        \begin{table*}[htpb!]
            \centering
\caption{Parameters in the External Compton (EC) scenario for epochs MJD 54761 (Figs. \ref{fig:sed_mjd54761_ec_w_bulk} and \ref{fig:sed_mjd54761_ec_w_bulk_conical}), MJD 55098 (Fig. \ref{fig:sed_mjd55098_ec_plc}), MJD 56576 (Fig. \ref{fig:sed_mjd56576_ec_plc}) and MJD 57293 (Fig. \ref{fig:sed_mjd57293_ec_plc_no_bulk}).}

\begin{adjustbox}{width=1.0\textwidth, center}
    
\begin{tabular}{w{l}{4cm}rlrrrr}
    
\toprule
Epoch & &
        & MJD 54761
        & MJD 55098
        & MJD 56576
        & MJD 57293
\\ 
\midrule
Model & &
& \makecell[r]{EC-dominated \\ + BC\tablefootmark{a}} 
& \makecell[r]{EC-dominated}
& \makecell[r]{EC-dominated}
& \makecell[r]{EC-dominated}
\\
\midrule
\multicolumn{3}{l}{Geometrical parameters}
\\[0.2em]
~ Bulk Lorentz factor & $\Gamma$ &
            & \num{34.0}		  (\num{<1e-1})
            & \num{20.4}	(\num{5e-01})
            & \num{16.5}	(\num{<e-1})
            & \num{25}		 (\num{3})
    \\
~ Viewing angle  & $\Theta$  & \si{\degree}
            & \num{1.50}		     (\num{<1e-02})
            & \num{1.33}    	(\num{6e-02})
            & \num{1.07} 		(\num{<e-2})
            & \num{1.40}        (\num{8e-02})    
    \\
~ Opening angle  & $\theta$ & \si{\degree}
            & \num{3.0}
            & =
            & =  
            & =                 
    \\
~ Location of the emission region & $r$   & \si{pc}
        & \num{5.41}	      (\num{<1e-2})
        & \num{4.52}   
        & \num{4.78}          (\num{<e-02})
        & \num{4.60}          (\num{9e-2})       
    \\
~ Size of the emission region & $R$ & \si{pc}
            & \num{0.28}      
            & \num{0.24}     
            & \num{0.25}      
            & \num{0.24}     
    \\
~ Light crossing time &  $t^{\mathrm{obs}}_{\mathrm{var}}(R,\Gamma,\theta)$  & \si{day}
            & \num{17}      (\num{1})
            & \num{15}      (\num{1})
            & \num{19}      (\num{1})
            & \num{15}      (\num{3})
\\
\midrule
Magnetic field intensity & $B$ & \si{G}
        & \num{6.03e-2}		              (\num{<1e-04})
        & \num{2.9e-1}				      (\num{2e-02})
        & \num{5.00e-2} 				(\num{<e-04})
        & \num{7.7e-2}                    (\num{4e-03})
\\
\midrule
{Particle distribution} & & 
        \\[0.2em]
        ~ Minimum Lorentz factor & $\gamma_{\text{min}}$ &
            & \num{1.06}	(\num{<1e-02})
            & \num{1.10}	(\num{3e-02})
            & \num{1.06}	(\num{<e-02})
            & \num{1.6}      (\num{3e-01})
        \\
        ~ Maximum Lorentz factor & $\gamma_{\text{max}}$  &
            & \num{7.20e+05}		  (\num{<1e+03})
            & \num{7.6e+05} 	  (\num{6e+04})
            & \num{1.11e+05}	(\num{<e3})
            & \num{9.1e4}            (\num{2e+03})
        \\
        ~ Type & $n_{e^-}(E)$ &
            & PLC
            & =       
            & =          
            & =                   
        \\
        ~ Density & N & \si{cm^{-3}}
            & \num{3.39e+01}		  (\num{<1e-1})
            & \num{4.7e+00}	      (\num{3e-01})
            & \num{6.96e+01}    (\num{<e-1}	)
            & \num{5.7e+01}       (\num{1e+01}) 
        \\
        ~ Cutoff Lorentz factor & $\gamma_{\text{cutoff}}$ &
            & \num{4.70e+03}			(\num{<1e+01})
            & \num{3.6e3}				(\num{1e+02})
            & \num{6.6e+03}	 	     (\num{7e+02})
            & \num{4.0e+03} 		 (\num{2e+03})\\ 
        ~ Spectral slope & $p$ &
            & \num{2.05}		 (\num{<1e-02})
            & \num{2.35}		 (\num{1e-02})
            & \num{2.30}  		 (\num{<e-2})
            & \num{2.4}    	      (\num{1e-01})      
\\
\midrule
Accretion disk & & 
         \\[0.2em]
          ~ Black hole mass & $M_{\text{BH}}$ & \si{M_{\odot}} 
                & \num{5e8}
                & =
                & =
                & =
          \\
          ~ Accretion efficiency & $\eta$ &
            & \SI{8e-2}{}    
            & =
            & =
            & =
          \\
          ~ Disk inner radius & $R_{\text{disk, in}}$ & \si{R_{S}}
            & \num{3.0}
            & =
            & =
            & =
          \\
         ~ Disk outer radius & $R_{\text{disk, out}}$  & \si{R_{S}}
            & \num{5e2}
            & =
            & =
            & =
         \\
          ~ Disk luminosity & $L_{\text{disk}}$  & \si{erg \, s^{-1}}
            & \num{5.0e45}
            & = 
            & = 
            &  = 
         \\
         ~ Disk temperature & $T_{\text{disk}}$  & \si{K}
            & \num{5.96e+04}
            & = 
            & =    
            & =    
\\
\midrule
Disk torus (DT)& &
        \\[0.2em]
        ~ Temperature & $T_{\text{DT}}$   & \si{K}
            & \num{330}
            & =
            & =
            & =
        \\
~ \makecell[{{p{4cm}}}]{Fraction of disk luminosity reprocessed} & $\tau_{DT}$ &
            & \SI{0.1}{}
            & = 
            & = 
            & =           
\\
\midrule
\multicolumn{2}{l}{Broad Line Region (BLR)} &
        \\
        ~ Inner radius & $R_{\text{BLR, in}}$  & \si{pc}
            & \num{6.87e-2}
            & = 
            & =    
            & =       
        \\
        ~ Outer radius & $R_{\text{BLR, out}}$  & \si{pc}
            & \num{7.56e-2}
            & = 
            & =   
            & =             
        \\
~ \makecell[{{p{4cm}}}]{Fraction of disk luminosity reprocessed} & $\tau_{BLR}$ &
            & \SI{0.1}{}
            & = 
            & =   
            & =            
\\
\midrule
\multicolumn{2}{l}{Fit statistics} &
\\
    ~ degrees of freedom & d.o.f &
    & 14
    & 16
    & 18
    & 15
\\
    ~ chi-quared statistic  & $\chi^2$ &
       & 21.9
       & 16.0
       & 10
       & 10.1
\\
\bottomrule
\end{tabular}

\end{adjustbox}

\tablefoot{
    Uncertainties for the best-fit values were automatically obtained using the \texttt{HESSE} method of second derivatives and are indicated between parenthesis, parameters without them were frozen during the fit.  For uncertainties smaller than the third significant digit, an upper limit is given. The degrees of freedom and the $\chi^2$ statistic for the model fit are indicated in the last rows, the residuals are shown in the figures.
    \\
    \tablefoottext{a}{The model for this epoch includes an additional component which is independently modeled as Bulk Compton emission from the disk,  the reported values refer to the SSC/EC components alone, with the exclusion of the X-ray data. Two possible geometries where considered for the EC-dominated scenario, and their parameters can be found Table \ref{tb:sed_models_bc}.}
    }             \label{tb:sed_models_ec_plc}
        \end{table*}
    
        \begin{table*}[htbp!]
            \centering
\caption{Parameters in the SSC-dominated scenario scenario for epochs MJD 54761, MJD 55098, MJD 56576 and MJD 57293, corresponding to models shown in Figs. \ref{fig:sed_mjd54761_ssc_w_bulk},  \ref{fig:sed_mjd55098_ssc_plc} and \ref{fig:sed_mjd56576_ssc_plc}.}

\begin{tabular}{w{l}{4cm}rlrrrr} \toprule
    Epoch & &
    & MJD 54761
    & MJD 55098
    & MJD 56576
\\ 
    \midrule
    \multirow{2}{*}{Model components} & &
    &  \multirow{2}{2.5cm}{\raggedleft SSC-dominated + BC\tablefootmark{a}}
    &  \multirow{2}{2.5cm}{\raggedleft SSC-dominated}
    &  \multirow{2}{2.5cm}{\raggedleft SSC-dominated}
\\
    \\
    \midrule
    \multicolumn{3}{l}{Geometrical parameters}
    \\[0.2em]
~ Bulk Lorentz factor & $\Gamma$ &
    & 38.4			(\num{5})
    & 21.1			(\num{1})
    & 25.0    	    (\num{<e-2})
\\
~ Viewing angle  & $\Theta$ & \si{\degree}
    & \num{1.55}		(\num{1e-2})
    & \num{1.79}	 	(\num{<e-2})
    & \num{1.66}       (\num{<e-2})
\\
~ Opening angle  & $\theta$ & \si{\degree}
    & \num{3.0}
    & \num{1.5}
    & \num{1.5}
\\
~ Location of the emission region & $r$  & \si{pc}
    & \num{32.4}		 (\num{<e-1})
    & \num{16.2}    
    & \num{16.2} 	 	(\num{<e-1})
\\
~ Size of the emission region & $R$   & \si{pc}
    & \num{1.70}
    & \num{0.42}       
    & \num{0.42}         
\\
~ Light crossing time &  $t^{\mathrm{obs}}_{\mathrm{var}}(R,\Gamma,\theta)$  & \si{day}
    & \num{106}      (\num{20})
    & \num{33}      (\num{3})
    & \num{30}      (\num{1})
\\
    \midrule
    Magnetic field intensity & $B$   & \si{G}
    & \num{2.00e-03}		(\num{<e-5})
    & \num{6.74e-03}		(\num{7e-04})
    & \num{5.63e-03}   	 	(\num{<e-5})
\\
    \midrule
    {Particle distribution} & & 
    \\[0.2em]
    ~ Minimum Lorentz factor & $\gamma_{\text{min}}$ &
    & \num{4.38e+01}      (\num{<1e-1})
    & \num{1.10e+02} 	  (\num{<1}) 
    & \num{4.57e+01}      (\num{<e-1})
\\
    ~ Maximum Lorentz factor & $\gamma_{\text{max}}$ &
    & \num{9.12e+06}		(\num{4e4})
    & \num{8.71e+05}   	 	(\num{<e+03})
    & \num{7.30e+05}              (\num{<e3})
\\
    ~ Type & $n_{e^-}(E)$ &
    & PLC
    & PLC       
    & PLC          
\\
    ~ Density & N & \si{cm^{-3}}
    & \num{9.94e-02}	  (\num{3e-4})
    & \num{5.35e-01}	   (\num{<e-3})
    & \num{8.51e-01}   	  (\num{<e-3})
\\
    ~ Cutoff Lorentz factor & $\gamma_{\text{cutoff}}$ &
    & \num{1.50e+04} 		(\num{<e2})
    & \num{9.33e+03}	   	(\num{<e1})
    & \num{8.73e+03}	 	(\num{<e1})
\\ 
    ~ Spectral slope & $p$ &
    & \num{1.50}	 (\num{<e-2})
    & \num{1.64}	 (\num{<e-2})
    & \num{1.68}	 (\num{<e-2})
\\
\\
    \midrule
Accretion disk & & 
             \\[0.2em]
              ~ Black hole mass & $M_{\text{BH}}$ & \si{M_{\odot}}
                    & \num{5e8}
                    & =
                    & =
\\
              ~ Accretion efficiency & $\eta$ & 
                & \num{8e-2} 
                & =
                & =
\\
              ~ Disk inner radius & $R_{\text{disk, in}}$ & \si{R_S}
                & \num{3.0}
                & =
                & =
\\
             ~ Disk outer radius & $R_{\text{disk, out}}$ & \si{R_S}
                & \num{5e2}
                & =
                & =
\\
              ~ Disk luminosity & $L_{\text{disk}}$ & \si{erg \, s^{-1}}
                & \num{5.0e45}
                & = 
                & = 
\\
             ~ Disk temperature & $T_{\text{disk}}$  & \si{K}
                & \num{5.96e+04}
                & = 
                & =

    \\
    \midrule
Disk torus (DT)& &
            \\[0.2em]
            ~ Temperature & $T_{\text{DT}}$  (\si{K})
                & \num{330}
                & =
                & =
\\
~ \makecell[{{p{4cm}}}]{Fraction of disk luminosity reprocessed} & $\tau_{DT}$
                & \num{0.1}{}
                & = 
                & = 
\\
    \midrule
    \multicolumn{2}{l}{Broad Line Region (BLR)} & 
            \\
            ~ Inner radius & $R_{\text{BLR, in}}$ & \si{pc}
                & \num{6.87e-2}
                & = 
                & =    
\\
            ~ Outer radius & $R_{\text{BLR, out}}$  & \si{pc}
                & \num{7.56e-2}
                & = 
                & =   
\\
~ \makecell[{{p{4cm}}}]{Fraction of disk luminosity reprocessed} & $\tau_{BLR}$ &
                & \num{0.1}{}
                & = 
                & =   
\\
    \midrule
    \multicolumn{2}{l}{Fit statistics} &
    \\
        ~ degrees of freedom & d.o.f &
        & 20
        & 16
        & 10
    \\
        ~ chi-quared statistic  & $\chi^2$ &
        & 44.4
        & 6.9
        & 7.2
    \\
    \bottomrule
\end{tabular}

\tablefoot{
    Uncertainties for the best-fit values were automatically obtained using the \texttt{HESSE} method of second derivatives and are indicated between parenthesis, parameters without them were frozen during the fit.  For uncertainties smaller than the third significant digit, an upper limit is given. The degrees of freedom and the $\chi^2$ statistic for the model fit are indicated in the last rows, the residuals are indicated in the figures.
    \\
    \tablefoottext{a}{The model for this epoch includes an additional componed which is independly modeled as Bulk Compton emission from the disk. See Table \ref{tb:sed_models_bc} for the parameters of the BC emission model.}
}             \label{tb:sed_models_ssc_plc}
        \end{table*}

        \begin{figure}[hptb!]
            \centering
            \includegraphics[width=1\linewidth]{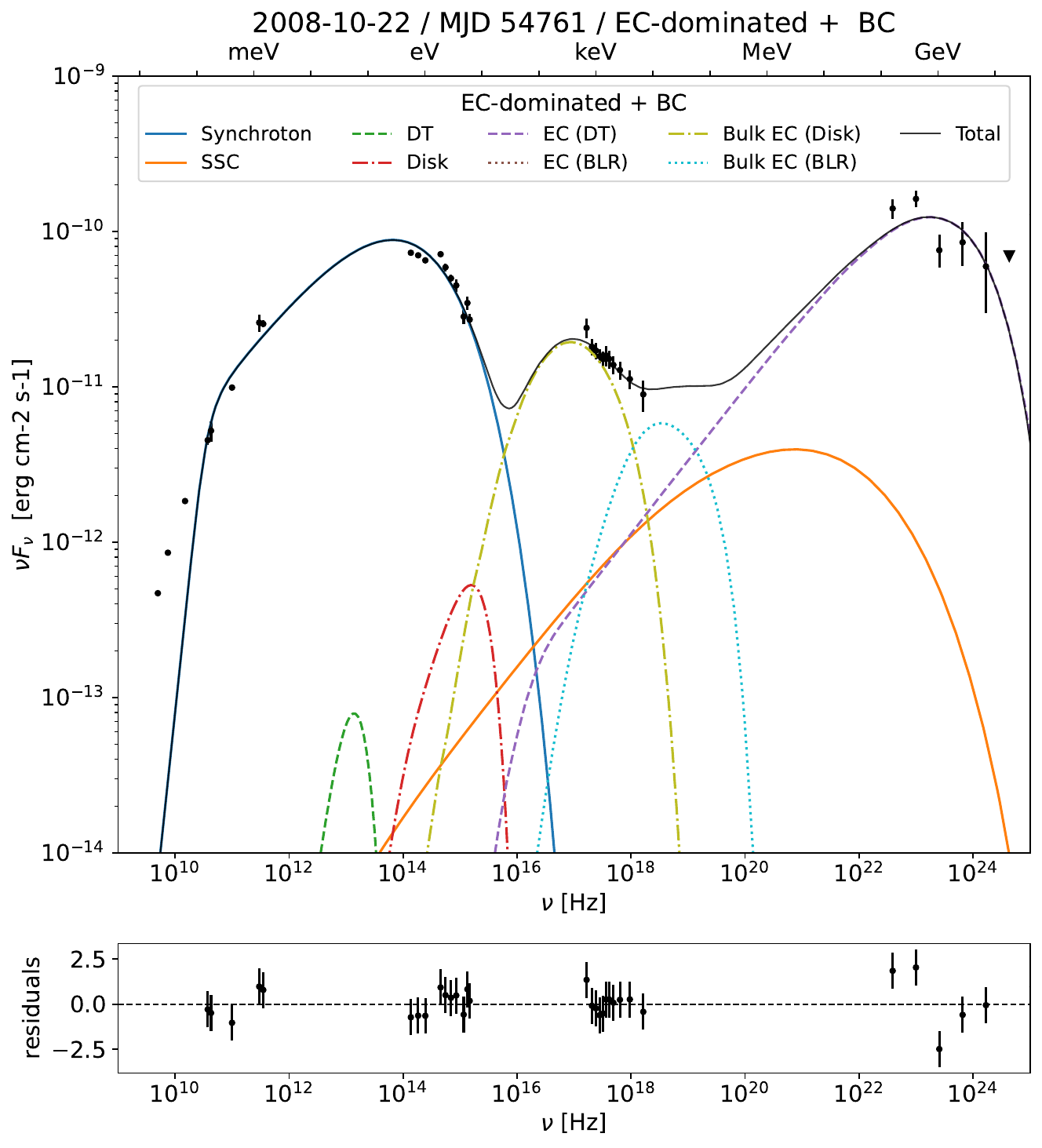}
            \caption{SED model for epoch MJD 54761. The model includes the usual synchrotron plus SSC components, but the high energy bump is dominated by EC emission from a dusty torus (Table \ref{tb:sed_models_ec_plc}). The X-ray bump is modeled by bulk Compton emission from the Disk by a secondary particle distribution much closer to the central engine (Table \ref{tb:sed_models_bc}), consistent with the much lower correlation and higher delays in the DCFs between X-ray and the other bands (Fig. \ref{fig:corr_I_compChart_tp}).}
            \label{fig:sed_mjd54761_ec_w_bulk}
        \end{figure}
    
        \begin{figure}[hptb!]
            \centering
            \includegraphics[width=1\linewidth]{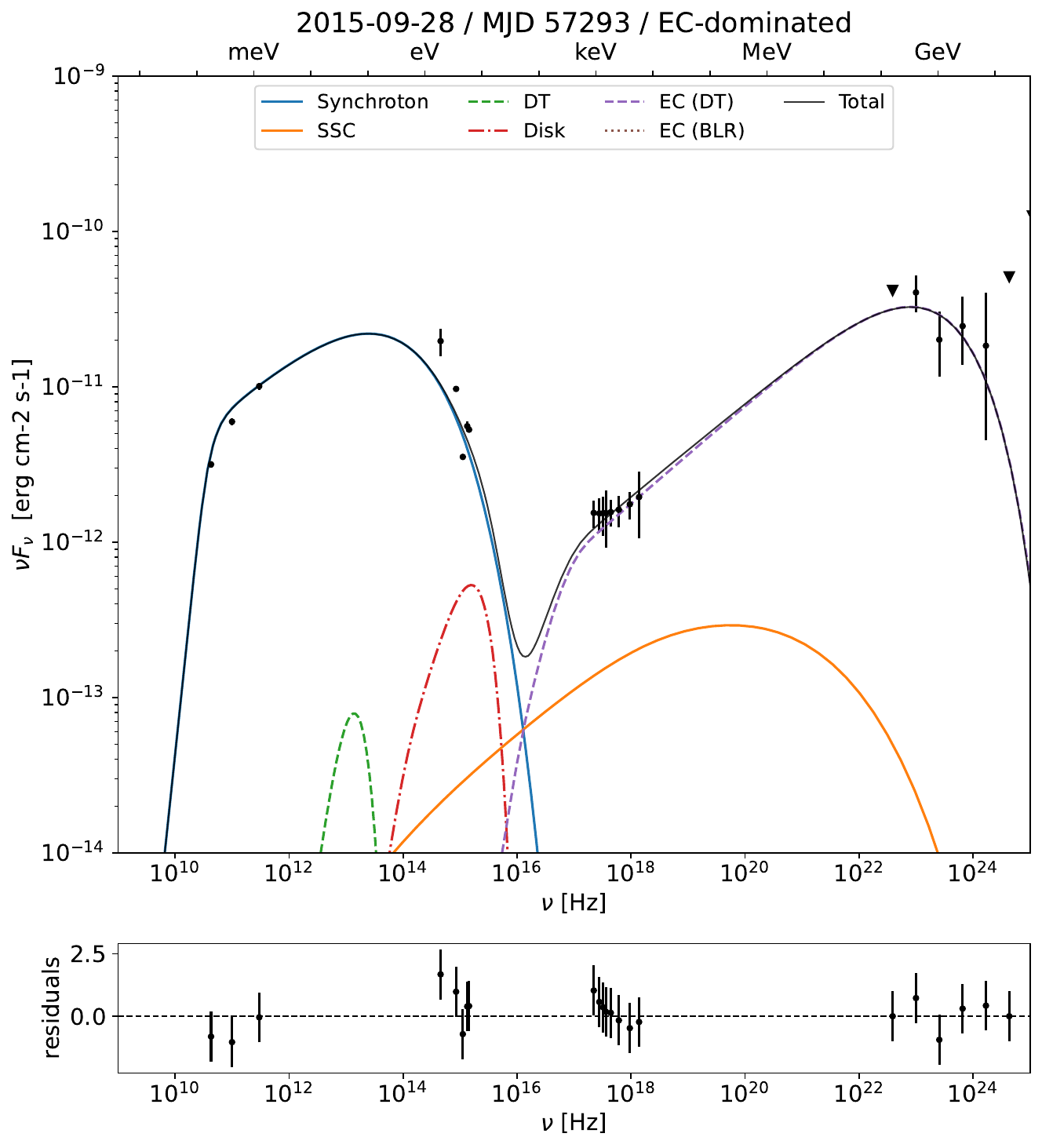}
            \caption{SED model for epoch MJD 57293 in the EC-dominated scenario (Table \ref{tb:sed_models_ssc_plc}). Unlike the models for the older flaring epoch 54761 (Figs. \ref{fig:sed_mjd54761_ec_w_bulk}, \ref{fig:sed_mjd54761_ec_w_bulk_conical}), this model does not include a bulk Compton component and can be explained with only the usual SSC+EC components. The source exhibits however an important softening the X-ray spectrum that could be explained also by bulk Compton emission.}
            \label{fig:sed_mjd57293_ec_plc_no_bulk}
        \end{figure}

    \section{Discussion and conclusions}
        
        We have presented new and updated multi-wavelength photometric and polarimetric data of  AO 0235+164, all across the spectrum from radio cm and mm wavelengths up to gamma-ray energies. The analysis of the correlations have shown that the emission at different wavelengths is statistically correlated, linking their emission mechanisms, with the notable exception of the X-ray band. 
        
        We have analyzed and shown the compatibility between the positions of the peaks of the different correlations, strengthening their interpretation as the delay between emissions. In this context, we have also shown that the obtained delays are compatible with the proposed emission mechanisms:  from mm to optical wavelengths, the delays agree with what it is to be expected for synchrotron emission. 
        
        In addition, we have seen that indeed also the $\gamma$-ray light curve is correlated with the mm and R-band emissions, which is to be expected if the dominating emission mechanism is SSC or EC.
Furthermore, the $\gamma$-ray subflares seem to be related to the appearance of identifiable VLBI components.

On the other hand, we have not found a significant correlation between the X-ray light curve and the rest of the bands. This is explained by the presence of the X-ray bump in the SED. This bump can not be accounted for by a closely correlated emission (SSC or EC) with the rest of the bands. Instead, it is proposed that it corresponds to bulk-Compton emission from a different population of particles. The large obtained delays imply that this emitting zone is separated by a large distance from the main emission component, and this is further confirmed by the results from SED modeling.
             
        Understanding how our observational data and results fit in the current landscape of existing blazar models is a difficult task. The rebrightening of knot features, which could be explained by successive recollimation shocks with the jet, and the difference in Doppler factor and speed between different components, which could be explained by different energies of a shock wave, points toward a shock-in-jet model.
The observed post-maximum subflares in \si{3}{mm} and $\gamma$-ray can be explained by less energetic recollimation of the same -dulled- shockwave-, analogously to the rebrightning of knot features farther from the jet as seen in the VLBA images, they even appear to be more or less simultaneous.
The observed longer duration of the flare in mm wavelengths is explained in this model by the longer cooling of synchrotron electrons. This smears out the peak in the correlation and shifts the correlation shape to show a delay of mm emission.

        The question about whether SSC or EC dominates the high energy bump does not have a clear, definite answer. EC-dominated SED models seem to be favored by literature (\cite{Ackermann:2012}, \cite{Dreyer:2021}). However, as we present in this paper, SSC-dominated models are also possible, as shown in section \ref{sec:seds}. It is generally easier and more common to produce a fit with dominant EC, however the model is harder to explain physically, and the obtained delays in correlation analysis and the results from VLBI observations favor SSC-dominated models. 
        
        The delays between signals are not directly interpretable as the relative time at which emissions at different wavelengths start, this interpretation would be valid only if the signals had the same shape but were shifted with respect to each other, which is not the case. But the correlation between R and $\gamma$ show a clear peak whose position is ${\tau_p}^{R,\gamma}$ of $\SI{2}{days}$, which corresponds to a distance of less than $\SI{1}{pc}$ after accounting for relativistic effects. Meanwhile, the large delay obtained between R and X-ray place the emission regions at tens of parsecs away, which nicely fits the obtained distances in the SSC scenario where the X-ray is produced by bulk-Compton emission.

        The results from the kinematic analysis of VLBI components show that the \SI{43}{GHz} core is located at distances from \SI{12}{pc} to \SI{17}{pc} downstream from the the central BH assuming a conical jet geometry. The best-fit distances obtained in SSC-models (Table \ref{tb:sed_models_ssc_plc}) are in better agreement with the ones obtained from the VLBI kinematic analysis, and in any case, since the SSC emission is less dependent on the distance to the BH, other distances are easier to accommodate; which is not the case in the EC-scenario.
        
        Scenarios where the $\gamma$-emitting zone is close to the central BH are ruled-out by the \emph{long-term and highly significant} correlation (Fig. \ref{fig:2020_ALL_corrs_pol}) between $\gamma$, R and mm light curves, since the emissions must be close enough and from analysis of VLBI images we know this is more than ten parsecs away from the central engine. SED models also help us discard these scenarios.

        The presence of IC flares after the synchrotron flares has already ended, such as some of those between the 2008 and 2015 flares, is also an indicator of SSC (\cite{Sokolov:2004}). They can be explained by the time-delays and crossing times, specially for small viewing angles such as AO 0235+164, but not in a EC scenario. Also the observed stronger variability in $\gamma$ rays with respect to  low energies is harder to explain in the EC scenario, where there is not a reasonable source of increased variability.
        
        A good test to determine whether the emission is SSC or EC might be polarization of the gamma-rays. EC is not expected to have significant polarization, while SSC is expected to have a polarization degree about half of the corresponding synchrotron emission. While X-ray polarization is already being measured by some instruments (IXPE), gamma-ray polarization is still not possible, although recent technological development open the possibility in the next decade.

    \section*{Acknowledgments}
    The IAA-CSIC team acknowledges financial support from the Spanish "Ministerio de Ciencia e Innovación" (MCIN/AEI/ 10.13039/501100011033) through the Center of Excellence Severo Ochoa award for the Instituto de Astrofísica de Andalucía-CSIC (CEX2021-001131-S), and through grants PID2019-107847RB-C44 and PID2022-139117NB-C44.
This research has made use of the NASA/IPAC Extragalactic Database (NED),
    which is operated by the Jet Propulsion Laboratory, California Institute of Technology,
    under contract with the National Aeronautics and Space Administration.
JYK was supported for this research by the National Research Foundation of Korea (NRF) grant funded by the Korean government (Ministry of Science and ICT; grant no. 2022R1C1C1005255).
IRAM is supported by INSU/CNRS (France), MPG (Germany) and IGN (Spain).
The VLBA is an instrument of the National Radio Astronomy Observatory, USA. The National Radio Astronomy Observatory is a facility of the National Science Foundation operated under cooperative agreement by Associated Universities, Inc.
    This study was based (in part) on observations conducted using the 1.8 m Perkins Telescope Observatory (PTO) in Arizona (USA), which is owned and operated by Boston University.
The BU group was supported in part by U.S. National Science Foundation grant AST-2108622, and NASA Fermi GI grants 80NSSC20K1567 and 80NSSC22K1571.

\bibliography{includes/citations} 
    
\begin{appendix}

\section{Swift Observations} \label{appendix:swift}

The {\em Neil Gehrels Swift observatory} satellite \citep{gehrels04} carried out 195 observations of AO 0235$+$164 between 2005 June 28 (MJD 53549) and 2016 February 11 (MJD 57429). The observations were performed with all three instruments onboard: the X-ray Telescope \citep[XRT;][0.2--10.0 keV]{burrows05}, the Ultraviolet/Optical Telescope \citep[UVOT;][170--600 nm]{roming05}, and the Burst Alert Telescope \citep[BAT;][15--150 keV]{barthelmy05}.

All XRT observations were performed in photon counting mode \citep[for a description of XRT read-out modes, see][]{hill04}. The XRT spectra were generated with the {\em Swift}-XRT data product generator tool at the UK Swift Science Data Centre\footnote{http://www.swift.ac.uk/user\_objects} \citep[for details, see][]{evans09}. Spectra having count rates higher than 0.5 counts s$^{-1}$ may be affected by pile-up. To correct for this effect, the central region of the image has been excluded, and the source image has been extracted with an annular extraction region with an inner radius that depends on the level of pile-up \citep[see e.g.,][]{moretti05}. We used the spectral redistribution matrices in the Calibration database maintained by HEASARC. The X-ray spectral analysis was performed using the \texttt{XSPEC 12.13.0c} software package \citep{arnaud96}. Data were grouped for having at least 20 counts per bins with \texttt{grppha} and the chi square statistics is used. All XRT spectra are fitted with an absorbed log-parabola model, except for cases with low number of counts, and a HI column density fixed to 2.8$\times$10$^{21}$ cm$^{-2}$ for taking into account the absorption effects of both our own Galaxy and an intervening $z$ = 0.524 system \citep[see e.g.][]{Madejski:1996}.

The hard X-ray flux of this source is usually below the sensitivity of the BAT instrument for daily short exposures. Moreover, the source is not included in the Swift-BAT 157-month catalogue\footnote{https://swift.gsfc.nasa.gov/results/bs157mon/}.

During the {\em Swift} pointings, the UVOT instrument observed the sources in its optical ($v$, $b$, and $u$) and UV ($w1$, $m2$, and $w2$) photometric bands \citep{poole08,breeveld10}. The UVOT data in all filters were analysed with the \texttt{uvotimsum}  and \texttt{uvotmaghist} tasks and the 20201215 CALDB-UVOTA release. Source counts were extracted from a circular region of 5 arcsec radius centered on the source, while background counts were derived from a circular region with a 20 arcsec radius in a nearby source-free region.

\end{appendix}

\end{document}